\documentclass[aps,superscriptaddress]{revtex4}

\usepackage{epsfig}
\usepackage{graphicx}
\usepackage{amsmath}
\usepackage{bm}
\usepackage{amssymb}
\usepackage{color}

\usepackage[export]{adjustbox} 

\newcommand{\cE}{\widetilde E}

\newcommand{\be}{\begin{equation}}
\newcommand{\ee}{\end{equation}}
\newcommand{\ba}{\begin{eqnarray}}
\newcommand{\ea}{\end{eqnarray}}
\newcommand{\bea}{\begin{eqnarray}}
\newcommand{\eea}{\end{eqnarray}}

\newcommand{\mut}{\tilde\mu}
\newcommand{\mub}{\mu_{B}}
\newcommand{\gt}{g}
\newcommand{\bk}{\mathbf k}
\newcommand{\bp}{\mathbf p}

\newcommand{\vt}{\tilde v}
\newcommand{\Kt}{\widetilde K}

\DeclareMathOperator{\sgn}{sgn}

\newcommand{\refer}[1]{(\ref{#1})}
\newcommand{\dd}{d}

\begin{document}

\title{Broken symmetry states in bilayer graphene in electric and in-plane magnetic fields}
\date{\today}

\author{Junji Jia}
\affiliation{School of Physics and Technology, and MOE Key Laboratory of Artificial Micro- and Nano-structures, Wuhan University, 430072, China}

\author{P. K. Pyatkovskiy}
\affiliation{Department of Physics and Astronomy, University of Manitoba, Winnipeg,
MB R3T 2N2, Canada}

\author{E. V. Gorbar}
\affiliation{Department of Physics, Taras Shevchenko National Kiev University, 03022, Kiev, Ukraine}
\affiliation{Bogolyubov Institute for Theoretical Physics, 03680, Kiev, Ukraine}

\author{V. P. Gusynin}
\affiliation{Bogolyubov Institute for Theoretical Physics, 03680, Kiev, Ukraine}

\begin{abstract}
Broken symmetry states in bilayer graphene in perpendicular electric $E_\perp$ and in-plane magnetic
$B_\parallel$ fields are studied in the presence of the dynamically screened long-range Coulomb
interaction and the symmetry-breaking contact four-fermion interactions. The integral gap equations are solved numerically, and it is shown that the momentum dependence of gaps is essential: It diminishes by an order of magnitude the gaps compared to the case of momentum-independent approximation,
and the obtained gap magnitudes are found to agree well with existing experimental values. We  derived
a phase diagram of bilayer graphene at the neutrality point in the plane $(B_\parallel,E_\perp)$ showing that the (canted) layer antiferromagnetic (LAF) state  remains a stable ground state of the system at
large $B_\parallel$. On the other hand, while  the LAF phase is realized at small values of $E_{\perp}$,  the quantum valley Hall (QVH) phase is the ground state of the system at values $E_{\perp}>E_{cr}(B_\parallel)$, where a critical value $E_{cr}(B_\parallel)$ increases with in-plane magnetic field $B_{||}$.
\end{abstract}

\maketitle
\section{Introduction}
\label{1}

The interlayer hopping $\gamma_1$ in Bernal stacked bilayer graphene modifies \cite{McC} the linear Dirac-type spectrum of charge carriers realized at high energy to the quadratic spectrum
at low energy. Quasiparticles in bilayer graphene are gapless and are characterized by the particle-hole
symmetric parabolic conduction and valence bands with massive chiral charge carriers, touching at two
$K$ and $K^{\prime}$ valley points. The crossover between the linear Dirac and quadratic dispersion takes
place at energy $E\simeq\gamma_1/4$.

The quadratic spectrum in bilayer graphene immediately implies [\onlinecite{Min}] that the electron-electron interaction should open a gap in the spectrum at the neutrality point in clean bilayer samples. The corresponding reasoning is straightforward. Since the electron density of states vanishes at the Dirac points in monolayer graphene, the condensate
of electron-hole pairs and the quasiparticle gap are formed only when the coupling constant exceeds a certain critical value \cite{GGMSh,Khveshchenko,Gamayun,Fertig,Gonzalez}. In this case the condensate
and gap acquire the exponential Berezinsky-Kosterlitz-Thouless-like behavior with respect to the
coupling constant. On the other hand, the density of states for the quadratic spectrum is nonzero and
a gap in bilayer graphene is generated for an arbitrary small interaction and has the exponential
Bardeen-Cooper-Schrieffer-like behavior for static screening, and the gap has a power-law scaling in
the coupling strength  for the dynamically screened Coulomb interaction \cite{Levitov,footnote}.
These theoretical considerations are confirmed by the experimental data, where no sign of an insulating state \cite{Mayorov} down to $1$\,K and gaps \cite{Weitz,Freitag,Velasco,Bao} of order $5$--$25$\,K are observed in monolayer and bilayer graphene, respectively, in the absence of external electromagnetic fields.

In this connection, we note that in the presence of a strong perpendicular magnetic field the kinetic
energy is quenched on the lowest Landau level and the quasiparticle band is completely flat. This leads
to the gap generation at any small coupling in monolayer graphene \cite{catalysis,GGMSh,MC} and enhancement of gaps in bilayer graphene.

The low-energy electron Hamiltonian with the Coulomb interaction in bilayer graphene possesses the approximate spin-valley SU(4) symmetry. This opens many interesting possibilities for the choice of
an order parameter. For example, in the absence of electromagnetic fields, the quantum anomalous Hall (QAH) \cite{Levitov-anomalous,Fzhang}, the quantum spin Hall (QSH), and the layer antiferromagnet states were
suggested as possible gapped ground states of bilayer graphene at the neutrality point (for a general discussion, see Ref.~[{\onlinecite{MacDonald}]). The QAH and QSH state gaps are due to the Haldane
mass \cite{Haldane} symmetric and antisymmetric in spin, respectively. The LAF state gap is described
by the Dirac mass antisymmetric in spin whereas the quantum valley Hall  state gap is described
by the Dirac mass symmetric in spin.

Experimentally, the broken symmetry states in bilayer graphene were studied in the presence
of a rather strong perpendicular magnetic field in
Refs.~[\onlinecite{Weitz,Martin,Freitag,Velasco,Feldman,Zhao,Kim,Elferen,Hunt}], where it was found
that the eightfold degeneracy in the zero-energy Landau level can be lifted completely, giving rise
to the quantum Hall states with filling factors $\nu=0,\pm 1,\pm 2, \pm 3$. These states
have been investigated theoretically in Refs.~[\onlinecite{Barlas,Abergel,Shizuya,Nakamura,GGM1,Levitov,Nandkishore1,Toke,Kharitonov1,GGJM,GGMSh-mixing}]
and a reasonable agreement with the experimental data  was found. However, the nature of the ground state of bilayer graphene in the absence or weak out-of-plane component of a magnetic field remains
a matter of debate.

Since the SU(4) symmetry is approximate in bilayer graphene, the role of the interaction
terms in the Hamiltonian which break this symmetry is crucially important. The valley-asymmetric interactions, which arise from the Coulomb interactions at the lattice scale or electron-phonon
interactions with the optical phonon modes, were considered in
Refs.~[\onlinecite{Lemonik,Vafek,Kharitonov,Kharitonov1,Kharitonov-AF}]. Taking these interactions
into account and using the experimental data as a guide, the canted antiferromagnetic (CAF) state
was suggested  as the ground state in both monolayer \cite{Kharitonov-mono} and bilayer graphene \cite{Kharitonov,Kharitonov1,Kharitonov-AF} at the neutrality point in a magnetic field.

The existing studies of the CAF state [\onlinecite{Kharitonov,Kharitonov1,Kharitonov-AF}] considered
a model  with only local four-fermion interactions where in the mean-field approximation the gap
equations are algebraic and generated gaps are constant. It is well known that the long-range Coulomb interaction does not permit constant gaps as solutions of integral gap equations, and momentum-dependent gaps have essentially smaller magnitude (see, for example, Ref.~\cite{Gamayun}). Therefore, it is important to investigate the ground state in bilayer graphene when both the long-range Coulomb interaction and the  SU(4) symmetry-breaking local interactions are present.
In addition, experimental studies of bilayer graphene in external fields of different orientation have become available recently \cite{Maher,Freitag2}. This provides the motivation for the study of the broken symmetry states in bilayer graphene at the neutrality point in the present paper, where we pay special attention to the role of the long-range Coulomb interaction as well as perpendicular electric and parallel magnetic fields on the broken symmetry states in bilayer graphene. Furthermore, we consider the case of a weak perpendicular magnetic field treating it as a parameter of the perturbative expansion.

The paper is organized as follows. We begin by presenting in Sec.~\ref{2} the model describing low-energy quasiparticle excitations in bilayer graphene in an external magnetic field interacting by means of the long-range Coulomb interaction and the local four-fermion interactions. The gap equations are derived in
Sec.~\ref{3} and analyzed in electric and in-plane magnetic fields in Sec.~\ref{secb0} in the
case where out-of-plane magnetic field $B_{\perp}$ is absent. Solutions in in-plane magnetic
field are found in Sec.~\ref{secb}. The linear and quadratic in $B_{\perp}$ corrections to the gap equation are considered in Sec.~\ref{5}. The results obtained in
the paper are summarized and discussed in Sec.~\ref{6}. In Appendix~\ref{append_k_dep_gaps},
the gap equations for the momentum-dependent generalized chemical potentials and gaps are derived. The
gap equations in the second order in $B_{\perp}$ are written down in Appendix~\ref{gap-eq-in-Bperp}.

\section{Model}
\label{2}

We utilize the same model for describing the low-energy electronic excitations as in Refs.~[\onlinecite{GGM1,GGJM,GGMSh-mixing}]. The free part
of the effective low-energy Hamiltonian
of bilayer graphene reads
\begin{equation}
H_0 = -\frac{1}{2m_*}\sum_{\xi,s}\int
d^2\mathbf{r}\,\Psi_{\xi s}^{\dagger}(\mathbf{r})\left( \begin{array}{cc} 0 & (\pi^{\dagger})^2\\
\pi^2 & 0\end{array} \right)\Psi_{\xi s}(\mathbf{r}), \label{free-Hamiltonian}
\end{equation}
where $\pi=\hat{p}_{x}+i\hat{p}_{y}$ and the canonical momentum $\hat{\mathbf{p}} = -i\hbar\bm{\nabla}
+ {e\mathbf{A}}/c$ includes the vector potential $\mathbf{A}$ corresponding to the component of an
external magnetic field $\mathbf{B}_{\perp }$ perpendicular to the bilayer planes.
The quasiparticle mass is $m_*= \gamma_1/2v_{F}^2\approx 0.054 m_e$, where $v_{F}\approx 8.0\times 10^{5}~\mbox{m/s}$ is the Fermi velocity, $\gamma_1 \approx 0.39~\mbox{eV}$, and $m_e$ is the mass of
the electron. The two-component spinor field $\Psi_{\xi s}$ carries the valley ($\xi=K,K^{\prime}$) and
spin ($s = \pm 1$) indices. We use the standard  convention \cite{McC}:
$\Psi_{K s}^T=(\psi_{KA_1}, \psi_{KB_2})_{s}$ for valley $K$ and
$\Psi_{K^{\prime}s}^T = (\psi_{K^{\prime} B_2}, \psi_{K^{\prime} A_1})_{s}$ for valley $K^{\prime}$.
Indices $A_1$ and $B_2$ label the corresponding $A$ and $B$ sublattices in the layers 1 (top) and 2 (bottom), respectively, which, according to the Bernal $(A_2-B_1)$ stacking, are relevant for the
low-energy dynamics.

The Zeeman and Coulomb interactions plus the top-bottom gates voltage imbalance $m_0$ (we denote
it $m_0$ because it corresponds to the time-reversal invariant bare Dirac mass) in bilayer graphene
are described as follows:
\begin{eqnarray}
H_{\rm int}=\int d^2\mathbf{r}\,\Psi^{\dagger}(\mathbf{r}) \left[\mu_B \bm{\sigma}\mathbf{B} +
m_0 \eta_3 \tau_3 \right]\Psi(\mathbf{r})+\frac{1}{2}\int d^2\mathbf{r}d^2\mathbf{r}^{\prime}\,
\Big\{V(\mathbf{r}-\mathbf{r}^{\prime})\,\rho(\mathbf{r})\rho(\mathbf{r}^{\prime}) +
2V_{\mbox{\scriptsize IL}}(\mathbf{r}-\mathbf{r}^{\prime})\rho_1(\mathbf{r})
\rho_2(\mathbf{r}^{\prime})\Big\}\,,
\label{interaction1}
\end{eqnarray}
where $\Psi$ combines the fields $\Psi_{K s}$ and $\Psi_{K' s}$ into an eight-component spinor.
Here $\mu_B$ is the Bohr magneton, $\mathbf{B}=\mathbf{B}_{\perp}+\mathbf{B}_\parallel$ is the total magnetic field with component $\mathbf{B}_\parallel$ parallel to the bilayer planes, $\mathbf{\sigma}_{i}$
are Pauli matrices in spin space, $\eta_3$ is the third Pauli matrix acting on the valley index of the fermion field, and $\tau_3$ is the diagonal Pauli matrix acting on the two components of the fields
$\Psi_{K s}$ and  $\Psi_{K^{\prime} s}$. Note that the presence of $\eta_3$ in the voltage imbalance
term is related to the different order of the $A_1$ and $B_2$ components in $\Psi_{K s}$ and $\Psi_{K^{\prime} s}$. The bias voltage between the top and bottom gates is related to the electric
field $E_{\perp}$ applied perpendicularly to the bilayer planes: $m_0=e E_{\perp }d /2$, where
$d=3.5\times 10^{-10} \,\mbox{m}$ is the distance between the layers. In our model, the in-plane component of $B_\parallel$ enters only through the Zeeman term, and we neglect its orbital effects due to finite $d$, which modify the electron spectrum \cite{Pershoguba,Roy3,Peeters,Falko} at energies smaller than the trigonal warping scale of about $1$\,meV \cite{McC} even at highest accessible fields.

The Coulomb interaction term $V(\mathbf{r})=e^2/(\kappa|\mathbf r|)$ in $H_{\rm int}$ is the bare intralayer potential whose
Fourier transform is given by $V(p)={2\pi e^2}/(\kappa p)$, where $\kappa$ is the dielectric constant.
The Fourier transform of the interaction $V_{\mbox{\scriptsize IL}}(\mathbf{r})$ equals
$V_{\mbox{\scriptsize IL}}(p)=2\pi e^2 (\mbox{e}^{-pd}-1)/(\kappa p)$. The interaction
$V_{\mbox{\scriptsize IL}}(\mathbf{r})=e^2/(\kappa\sqrt{\mathbf{r}^2+d^2})-V(\mathbf r)$ describes the $d$-dependent part of the interlayer electron interactions and unlike the Coulomb interaction is not invariant with respect to the spin-valley
SU(4) symmetry. Since $d$ is small in bilayer graphene, this interaction is weak. The two-dimensional
charge densities in the two layers are (the total charge density $\rho=\rho_1+\rho_2$)
\begin{eqnarray}
\label{density1}
\rho_1(\mathbf{r})=\Psi^{\dagger}(\mathbf{r}){\cal P}_1\Psi(\mathbf{r})\,,\quad\quad
\rho_2(\mathbf{r})=\Psi^{\dagger}(\mathbf{r}){\cal P}_2\Psi(\mathbf{r})\,,
\label{density}
\end{eqnarray}
where ${\cal P}_1=(1+\eta_3\tau_3)/2$ and ${\cal P}_2=(1-\eta_3\tau_3)/2$ are projectors on the
states in the layers 1 and 2, respectively. When the dynamical screening effects are taken into
account, the potentials $V(\mathbf{r})$ and $V_{\mbox{\scriptsize IL}}(\mathbf{r})$ are replaced
by effective interactions $V_{\rm eff}(t,\mathbf{r})$ and $V^{\rm eff}_{\mbox{\scriptsize IL}}
(t,\mathbf{r})$ which are no longer instantaneous.

If external electric and magnetic fields are absent and the interaction $V_{\mbox{\scriptsize IL}}$
is neglected, then the Hamiltonian $H = H_0 + H_{\rm int}$, with $H_0$ and $H_{\rm int}$ in
Eqs.~(\ref{free-Hamiltonian}) and (\ref{interaction1}), possesses the spin-valley SU(4) symmetry and
all the QAH, QVH, QSH, and LAF states discussed in the Introduction are degenerate in energy at the neutrality point. In order to qualify
these states, we write down their order parameters (condensates) in terms
of the valley-layer components of the spinor $\Psi$:
\ba
&&\mbox{QAH}:\quad\langle\Psi^\dagger\tau_3\Psi\rangle=
\langle\psi^{\dagger}_{KA_1s}\psi_{KA_1s}-\psi^{\dagger}_{K'A_1s}\psi_{K'A_1s}
-\psi^{\dagger}_{KB_2s}\psi_{KB_2s}+\psi^{\dagger}_{K'B_2s}\psi_{K'B_2s}\rangle,
\label{condensate-QAH}
\\
&&\mbox{QVH}: \quad \langle\Psi^\dagger \eta_3\tau_3\Psi\rangle=
 \langle\psi^{\dagger}_{KA_1s}\psi_{KA_1s}+\psi^{\dagger}_{K'A_1s}\psi_{K'A_1s}
 -\psi^{\dagger}_{KB_2s}\psi_{KB_2s}-\psi^{\dagger}_{K'B_2s}\psi_{K'B_2s}\rangle,\\
&&\mbox{QSH}:\quad \langle\Psi^\dagger \sigma_3\tau_3\Psi\rangle=
 \langle\psi^{\dagger}_{KA_1s}s\psi_{KA_1s}-\psi^{\dagger}_{K'A_1s}s\psi_{K'A_1s}
-\psi^{\dagger}_{KB_2s}s\psi_{KB_2s}+\psi^{\dagger}_{K'B_2s}s\psi_{K'B_2s}\rangle,\\
&&\mbox{LAF}:\quad \langle\Psi^\dagger \sigma_3\eta_3\tau_3\Psi\rangle=
\langle\psi^{\dagger}_{KA_1s}s\psi_{KA_1s}+\psi^{\dagger}_{K'A_1s}s\psi_{K'A_1s}
-\psi^{\dagger}_{KB_2s}s\psi_{KB_2s}-\psi^{\dagger}_{K'B_2s}s\psi_{K'B_2s}\rangle,
\ea
where the summation over the spin index is implied. The QAH state describes a state in which
the $K$ and $K'$ valleys have opposite layer polarizations leading to the quantum Hall
effect even in the absence of a magnetic field. For the QVH state, the layer polarization is the
same for both valleys, breaking thus an inversion symmetry; therefore, this state can be called also
a layer-polarized state. The QSH and LAF states at fixed spin have the same order parameter as
the QAH and QVH states and, unlike the latter states, flip the sign of order parameter for the
opposite direction of spin. Thus, the states QAH, QVH and QSH, LAF are symmetric and antisymmetric
in spin, respectively. It is important to emphasize that the QVH and LAF states do not have
topologically protected edge states and finite Hall conductivities, while the QAH and QSH states
possess topologically protected edge states leading to nonzero charge and spin Hall conductivities, respectively \cite{Fzhang,MacDonald}. The time-reversal symmetry is unbroken for the QVH, QSH and
broken for QAH, LAF states. Since at the neutrality point $\rho=\rho_1+\rho_2=0$, the negative
interaction term $V_{\mbox{\scriptsize IL}}$ in Hamiltonian (\ref{interaction1}) makes the layer-polarized
QVH state have larger energy than the other three QAH, QSH, and LAF states, which remain degenerate
in energy.

Clearly, for a sufficiently large electric field $E_{\perp}$ perpendicular to the planes of graphene,
the layer-polarized QVH state should be realized. The experiments performed in Refs.~[\onlinecite{Weitz,Freitag}] demonstrated a phase transition to another state as $E_\perp $
decreases. This eliminates the QVH state as a possible candidate for the ground state of bilayer
graphene at the neutrality point in the absence of external fields. On the other hand, the recent
experiment \cite{Maher} revealed a quantum phase transition at large in-plane magnetic field to
a state with the conductance of order $4e^2/\hbar$ consistent with the QSH state. This also excludes
the QSH state as the ground state of bilayer graphene in the absence of in-plane magnetic
field.
Since the QAH state has topologically protected edge states, hence nonzero conductance, the
experimental data in Ref.~[\onlinecite{Maher}] single out the insulating LAF phase as the ground
state of bilayer graphene in the absence  of external electric and in-plane magnetic fields. According to Refs.~[\onlinecite{Kharitonov,Kharitonov1}], the LAF state transforms in an out-of-plane magnetic
field into the CAF state once the Zeeman coupling is taken into account. Moreover, for
$\mathbf{B}_{\perp}\ne 0$, the CAF state continuously crosses over into the  QSH state as in-plane
magnetic field increases.

Since the QAH, QSH, and LAF states are degenerate in energy for the Hamiltonian $H=H_0+H_\text{int}$
with the long-range Coulomb interaction in the absence of external fields, new terms breaking the
SU(4) symmetry should be added to the Hamiltonian in order to ensure that the LAF is the ground
state of bilayer graphene. To provide this, the following local four-fermion
interaction terms allowed by the symmetry of the bilayer lattice were added to the Hamiltonian of
the model in Refs.~\cite{Lemonik,Kharitonov,Kharitonov1}:
\begin{equation}
H_{\rm asym}=\frac{2\pi\hbar^2}{m_*}\sum_{\alpha,\beta=0}^3\gt_{\alpha\beta}\int d^2\mathbf{r}\,\left[\Psi^{\dagger}
(\mathbf{r})\eta_\alpha\tau_\beta\Psi(\mathbf{r})\right]^2,
\label{asym-interaction}
\end{equation}
where $\eta_j$ are the Pauli matrices acting on the valley degree of freedom of the fermion field
and we set the SU(4)-symmetric coupling constant $g_{00}$ of the local Coulomb interaction
to zero. As argued in Refs.~[\onlinecite{Kharitonov-mono,Kharitonov,Kharitonov-AF}], the U(4)-asymmetric interactions (\ref{asym-interaction}) arise actually from the Coulomb interaction at the lattice scale $a\approx2.46$\,{\AA} or electron-phonon interactions with the optical phonon modes, therefore, they
can be assumed to be local in the effective low-energy model. There are generically eight independent dimensionless coupling constants $g_{\perp\perp}\equiv\gt_{11}=\gt_{12}=\gt_{21}=\gt_{22}$,
$g_{\perp0}\equiv\gt_{10}=\gt_{20}$,
$g_{0\perp}\equiv\gt_{01}=\gt_{02}$,
$g_{\perp z}\equiv\gt_{13}=\gt_{23}$,
$g_{z\perp}\equiv\gt_{31}=\gt_{32}$,
$g_{zz}\equiv\gt_{33}$,
$g_{z0}\equiv\gt_{30}$, and
$g_{0z}\equiv\gt_{03}$,
whose bare values should be less or order of the dimensionless strength of the Coulomb interaction
at the lattice scale $e^2a m_*/\hbar^2\approx0.25$. As we will see, not all local couplings
$g_{\alpha\beta}$ are relevant for determining the phase diagram of the system, and only certain
combinations of them are important.

\section{Gap equation}
\label{3}

The Schwinger--Dyson (SD) or gap equation for the quasiparticle Green's function (propagator) in
the Hartree-Fock approximation reads \cite{GGM1,GGMSh-mixing}
\begin{eqnarray}
G^{-1}(x,y) &=&S^{-1}(x,y) -(\mu_B\bm{\sigma}\mathbf{B}+m_0\eta_3\tau_3)\,\delta^3(x-y)
- iG(x,y) V_{\rm eff} (x-y) \nonumber  \\
&-& i\left[{\cal P}_1G(x,y) {\cal P}_2+{\cal P}_2G(x,y) {\cal P}_1\right]
{V}^{\rm eff}_{\rm IL}(x-y) - i\frac{4\pi\hbar^2}{m_*}
\sum_{\alpha,\beta=0}^3\gt_{\alpha\beta}\eta_\alpha\tau_\beta
G(x,x)\eta_\alpha\tau_\beta\delta^3(x-y)\label{SD-equation}  \\
&-&\frac{i}{2}\left[{\cal P}_1-{\cal P}_2\right]\,\mbox{tr}\big\{\left({\cal P}_1-
{\cal P}_2\right)G(x,x)\big\}
V_{\rm IL}(0)\,\delta^3(x-y)+i\frac{4\pi\hbar^2}{m_*}\sum_{\alpha,\beta=0}^3
\gt_{\alpha\beta}\eta_\alpha\tau_\beta\mbox{tr}\big\{\eta_\alpha\tau_\beta G(x,x)\big\}\,
\delta^3(x-y),\nonumber
\end{eqnarray}
where $x=(t,\mathbf{x})$, $G(x,y)=\hbar^{-1}\langle0|\Psi(x)\Psi^\dagger(y)|0\rangle$ is the full propagator, $S(x,y)$ is the free propagator in the theory with $m_0=\mu_B|\mathbf{B}|=0$,
and $V_{\rm IL}(0)=-2\pi e^2d/\kappa$ is the Fourier transform of the interlayer interaction
$V_{\rm IL}(\mathbf{r})$ at zero momentum. Note also that due to the overall neutrality of the
system, we dropped all Hartree terms proportional to $\mbox{tr}[G(x,x)]$ (i.e., the charge density)
in the gap equation. The explicit form of the interlayer potential in momentum space
$V^{\rm eff}_{\rm IL}(\omega,p)$ can be found in the Appendix of the second
paper in Ref.~[{\onlinecite{GGM1}]. Here we do not need it: Due to the presence of the projectors
${\cal P}_1$ and ${\cal P}_2$ in the second line of Eq.~(\ref{SD-equation}), the corresponding Fock
term does not contribute to the final form of the gap equation if the Green's functions are diagonal
in the valley space. As to the effective interaction $V_{\rm eff}$, in  momentum space it reads
\begin{eqnarray}
V_{\rm eff}(\omega,p)=\frac{2\pi e^2}{\kappa}\frac{1}{\displaystyle p+\frac{\pi e^2}{\kappa}
\Pi(\omega,p,\Delta_{\alpha\beta})}\,,
\label{veff}
\end{eqnarray}
where $\Pi(\omega,p,\Delta_{\alpha\beta})$ is the polarization function. The one-loop polarization
function $\Pi(\omega,p,\Delta_{\alpha\beta})$ as an integral over momentum is given in Ref.~[\onlinecite{Levitov-anomalous}]. In this work, we will set quasiparticle gaps $\Delta_{\alpha\beta}$ to zero in the polarization function because it weakly depends on gaps.
In this case, the polarization function is given by \cite{Levitov}:
\begin{equation}
\Pi(\omega,k)=\frac{4m_*}{\pi\hbar^2}\,P\left(\frac{\omega}{k}\right),\quad\quad P(z)=\ln\biggl(\frac{4+4z^2}{1+4z^2}\biggr)+\frac{2\arctan(z)
-\arctan(2z)}z.
\label{polarization}
\end{equation}
It is convenient to define the full quasiparticle propagator $G$ through the self-energy $\Sigma$ as follows:
\begin{equation}
G^{-1}(x,z)=S^{-1}(x,z)+\Sigma(x,z)\,,
\label{self-energy}
\end{equation}
where $S^{-1}(x,z)$ is the free inverse propagator. Then the Schwinger--Dyson equation in terms of
the self-energy takes the following form:
\begin{eqnarray}
&&\Sigma(x,y)=-(\mu_B\bm{\sigma}\mathbf{B}+m_0\eta_3\tau_3)\,\delta^3(x-y)-i\,G(x,y)\,
V_{\rm eff}(x-y)-i\frac{4\pi\hbar^2}{m_*}\sum_{\alpha,\beta=0}^3\gt_{\alpha\beta}\eta_\alpha\tau_\beta G(x,x)\eta_\alpha\tau_\beta\delta^3(x-y)
\nonumber\\
&&-i\left[{\cal P}_1\,G(x,y)\,{\cal P}_2+{\cal P}_2\,G(x,y)\,{\cal P}_1\right]
V^{\rm eff}_{\mbox{\scriptsize IL}}(x-y)-\frac{i}{2}\left[{\cal P}_1-{\cal P}_2\right]\,
\mbox{tr}\big\{\left({\cal P}_1-{\cal P}_2\right)G(x,x)\big\}\,
V_{\mbox{\scriptsize IL}}(0)\,\delta^3(x-y)\nonumber\\
&&+i\frac{4\pi\hbar^2}{m_*}\sum_{\alpha,\beta=0}^3\gt_{\alpha\beta}\eta_\alpha\tau_\beta\mbox{tr}
\big\{\eta_\alpha\tau_\beta G(x,x)\big\}\,
\delta^3(x-y).
\label{SD-equation-self-energy}
\end{eqnarray}
Although the quasiparticle propagator is not translation invariant when an out-of-plane magnetic
field is present, it can be written in the form of the product of the non-translation-invariant
Schwinger phase $\Phi(x,y)=-(e/\hbar c)\mathbf{x}\mathbf{A}(\mathbf{y})$ in the symmetric gauge
$\mathbf{A}=(-B_{\perp}x_2/2,B_{\perp}x_1/2)$ and a translation invariant function
$$
G(x,y)=e^{i\Phi(x,y)}\tilde{G}(x-y)\,.
$$
Then in terms of translation invariant self-energy $\tilde{\Sigma}(x-y)=\mbox{exp}[-i\Phi(x,y)]
\Sigma(x,y)$ and propagator $\tilde{G}(x-y)$, Eq.~(\ref{SD-equation-self-energy}) takes the
following form in momentum space:
\begin{eqnarray}
\tilde{\Sigma}(\Omega,\mathbf{p})&=&-\mu_B\bm{\sigma}\mathbf{B}-m_0\eta_3\tau_3
-i\int\frac{d\omega d^2{k}}{(2\pi)^3}\,\tilde{G}(\omega,\mathbf{k})\,
V_{\mbox{\scriptsize eff}}(\Omega-\omega,\mathbf{p}-\mathbf{k})\nonumber\\
&-& i\int\frac{d\omega d^2{k}}{(2\pi)^3}\left(\,{\cal P}_1\tilde{G}(\omega,\mathbf{k})
{\cal P}_2+{\cal P}_2\tilde{G}(\omega,\mathbf{k}){\cal P}_1\right)
V_{\mbox{\scriptsize IL}}
(\Omega-\omega,\mathbf{p}-\mathbf{k})-i\frac{4\pi\hbar^2}{m_*}\sum_{\alpha,\beta=0}^3
\gt_{\alpha\beta}\int\frac{d\omega d^2{k}}{(2\pi)^3}\,\eta_\alpha\tau_\beta
\tilde{G}(\omega,\mathbf{k})\eta_\alpha\tau_\beta
\nonumber\\
&-&\frac{i}{2}\,\int\frac{d\omega d^2{k}}{(2\pi)^3}\left({\cal P}_1-{\cal P}_2\right)
\mbox{tr}\,[({\cal P}_1-{\cal P}_2)\tilde{G}(\omega,\mathbf{k})\,]\,
V_{\mbox{\scriptsize IL}}(0)+i\frac{4\pi\hbar^2}{m_*}\sum_{\alpha,\beta=0}^3\gt_{\alpha\beta}
\int\frac{d\omega d^2{k}}{(2\pi)^3}\,
\eta_\alpha\tau_\beta\mbox{tr}\big\{\eta_\alpha\tau_\beta\tilde{G}(\omega,\mathbf{k})\big\}\,.
\label{SD-equation-momentum}
\end{eqnarray}

Finally, in order to finish the setup of our problem, we should select an ans\"atz for the quasiparticle self-energy $\tilde{\Sigma}$. Since we consider the out-of-plane magnetic field $\mathbf{B}_{\perp}$ in perturbation theory, we use the following ans\"atz for the self-energy up to the second order in $\mathbf{B}_{\perp}$:
\begin{align}
\tilde{\Sigma}(\Omega,p)&=-\Delta_{\alpha\beta}(\Omega,p)\eta_{\alpha}\sigma_{\beta}\tau_3
-\mu_{\alpha\beta}(\Omega,p)\eta_{\alpha}\sigma_{\beta}
-\mu^{(1)}_{\alpha\beta}(\Omega,p)\eta_{\alpha}\sigma_{\beta}B_\perp
\nonumber \\
&\quad-\Delta^{(1)}_{\alpha\beta}(\Omega,p)\eta_{\alpha}\sigma_{\beta}\tau_3\,B_{\perp}
-\mu^{(2)}_{\alpha\beta}(\Omega,p)\eta_{\alpha}\sigma_{\beta}B_\perp^2
-\Delta^{(2)}_{\alpha\beta}(\Omega,p)\eta_{\alpha}\sigma_{\beta}\tau_3\,B_{\perp}^2.
\label{ansatz}
\end{align}
The symmetry-breaking quantities $\mu_{\alpha\beta}(p)$ and $\Delta_{\alpha\beta}(p)$ are
related to the corresponding order parameters through the following relationship:
\begin{equation}
\langle\Psi^\dagger{\cal O}^{\alpha\beta}\Psi\rangle=-\hbar{\,\rm tr}[{\cal O}^{\alpha\beta} G(x,x)],
\quad\langle\Psi^\dagger{\cal \tilde{O}}^{\alpha\beta}\Psi\rangle=-\hbar{\,\rm tr}
[{\cal \tilde{O}}^{\alpha\beta} G(x,x)],
\label{order_parameters}
\end{equation}
where ${\cal O}^{\alpha\beta}=\eta_\alpha\sigma_\beta$,
${\cal \tilde{O}}^{\alpha\beta}=\eta_\alpha\sigma_\beta\tau_3$, and the trace is taken over the sublattice, valley, and spin indices.

Equation~(\ref{SD-equation-momentum}) admits, in general, many solutions. In order to select the solution
which is the ground state of the system, we should calculate the energy density for each of these
states, which is given by \cite{GGMSh-mixing}
\begin{equation}
{\cal E}=\frac{i}{2}\int \frac{d\omega d^2p}{(2\pi)^3}\,\mbox{tr}\,
\left[\,\left(\,-\omega-\mu_B\bm{\sigma}\mathbf{B}-m_0\eta_3\tau_3 + \frac{\hbar^2}{2m_*}
\left( \begin{array}{cc} 0 & D^- \\
D^+ & 0 \end{array} \right)\,\right)\tilde{G}(\omega,\mathbf{p})\,\right]\,\,-\,\,(\mu_{\alpha\beta}
\to 0, \Delta_{\alpha\beta} \to 0)\,,
\label{energy-density-momentum}
\end{equation}
where $D^{\pm}$ are given by Eq.~(\ref{quadratic}) in Appendix~\ref{gap-eq-in-Bperp}, and then determine the solution with
the lowest energy density.
For all phases (at $B_\perp=0$), we use in what follows the full electron propagator in the Minkowski space,
\begin{equation}
\tilde G(\omega,\bk)=\frac{1}{\omega+D_0+\tilde{\Sigma} +i\epsilon\sgn\omega},
\qquad
D_0=\frac{\hbar^2}{2m_*}
\begin{pmatrix}
0 & (k_x-ik_y)^2 \\ (k_x+ik_y)^2 & 0
\end{pmatrix},
\label{G_and_D0}
\end{equation}
which we write in the form
\begin{equation}
\tilde G(\omega,\bk)=\sum_j\frac{A_j(\bk)}{\omega-\cE_j(k)+i\epsilon\sgn\omega},
\label{G_gen}
\end{equation}
where $\cE_j(k)$ are the energy dispersions (the index $j$ enumerates the branches of the spectrum)
and $A_j(\bk)$ are matrices  in the spin-valley-layer space.

Finally, since we consider bilayer graphene at the neutrality point, valid solutions should also
satisfy the charge neutrality condition
\be
n=i\int\frac{d\omega d^2k}{(2\pi)^3}\,\mbox{tr}\,\left[\,\tilde{G}(\omega, {\mathbf k})\,\right]=0,
\label{neutrality-cond}
\ee
where trace runs over spin, valley, and sublattice indices. All our
self-energy ans\"atze below satisfy the charge neutrality condition.

\section{Solutions in the absence of magnetic field}
\label{secb0}

In this section, we will solve the gap equations in electric field $E_{\perp}$ in the case where
external magnetic field is absent. In what follows, we set $\hbar=1$.

\subsection{The QAH, QSH,  QVH, and LAF states in the absence of electric field}

Let us discuss first the simplest case where the external electric field is absent and
study the QAH, QSH, QVH, and LAF states. According to Eqs.~(\ref{condensate-QAH}), (\ref{ansatz}), and (\ref{order_parameters}),
the order parameter $\langle\Psi^\dagger\tau_3\Psi\rangle$ of the QAH state is proportional to the
Haldane mass $\Delta_{00}$ and describes a charge density wave with an opposite sign
in the $K$ and $K^{\prime}$ valleys, which is odd under time reversal. The order parameter
$\langle \Psi^\dagger\sigma_j\tau_3\Psi\rangle\sim \Delta_{0j}$ ($j=1,2,3$) of the QSH state is antisymmetric in spin. The order parameter connected with the conventional Dirac mass $\langle
\Psi^\dagger \eta_3\tau_3\Psi\rangle\sim \Delta_{30}$ is the order parameter of the QVH state and
determines the charge-density imbalance between the two layers. The structure of this mass term
coincides with that of the voltage imbalance term $m_0$ between the top and bottom gates introduced
in Hamiltonian~(\ref{interaction1}) and, therefore, can be considered as a dynamical counterpart of
the latter. This mass term is even under time reversal. The Dirac mass term antisymmetric in spin
$\langle \Psi^\dagger \eta_3\sigma_j\tau_3\Psi\rangle \sim \Delta_{3j}$ ($j=1,2,3$) is
the order parameter of the LAF state. Since the direction of the QSH and LAF state gaps in the spin space
(subscript $j$ in $\Delta_{0j}$ and $\Delta_{3j}$) is completely arbitrary in the absence of an
external magnetic field, we choose for the sake of convenience the third direction $j=3$
in these gaps.

Let us derive the gap equations from the master gap equation~\refer{SD-equation-momentum} for the
QAH, QSH, QVH, and LAF states. Since there are no external electric and magnetic fields, the first two
terms on the right-hand side of Eq.~(\ref{SD-equation-momentum}) are absent.
Substituting the corresponding ans\"atz for the self-energy
into Eq.~\refer{SD-equation-momentum} allows us to establish
the gap equations for each state in terms of their gaps. We find it convenient to use the notations $\Delta$, $\Delta_z$, $m$, and $m_z$ for gaps $\Delta_{00}$, $\Delta_{03}$, $\Delta_{30}$, and $\Delta_{33}$,  respectively. Thus, we use $\Delta$ and $\Delta_z$ to describe the QAH and QSH gaps
with the Haldane-type masses and $m$ and $m_z$ to describe the QVH and LAF gaps with their Dirac-type
gaps.

We assume that the dependence of gap functions on the energy is rather weak so that we can approximate
these functions by their values at $\Omega=0$ and neglect the dependence of $V_{\mbox{\scriptsize eff}}$
on external energy, i.e., we approximate
$V_\text{eff}(\Omega-\omega,\bp-\bk)\approx V_{\mbox{\scriptsize eff}}(\omega,\mathbf{p}-\mathbf{k})$.
The momentum dependence of gaps generally results in the significant reduction in their
sizes compared to the case of momentum-independent gaps.

The gap equations for the QAH ($\delta=\Delta$), QSH ($\delta=\Delta_z$), LAF ($\delta=m_z$)
and QVH ($\delta=m$) phases have the form
\begin{equation}
\delta(p)=\delta_0+\int\limits^\Lambda\frac{d\omega d^2k}{(2\pi)^3}\frac{\delta(k)}{\omega^2+E_k^2
+\delta^2(k)} \left[V_{\mbox{\scriptsize eff}}
(\omega,\mathbf{p}-\mathbf{k})+\frac{4\pi}{m_*}g_\delta\right],
\label{gengeq1}
\end{equation}
where $E_k=k^2/(2m_*)$, the UV momentum integration cutoff $\sqrt{2m\Lambda}$ with $\Lambda=\gamma_1/4$
is used in our low-energy two-band model, the Wick's rotation $\omega\to i\omega$ has been made, and
we used the following notations for the different linear combinations of the local interaction constants
for the QAH, QSH, LAF, and QVH states:
\begin{align}
g_\Delta&=-2g_{z\perp}-7g_{0z}+2g_{\perp z}-4g_{\perp\perp}+2g_{\perp0}+g_{z0}-2g_{0\perp}+g_{zz}, \\
g_{\Delta_z}&=-2g_{z\perp}+g_{0z}+2g_{\perp z}-4g_{\perp\perp}+2g_{\perp0}+g_{z0}-2g_{0\perp}+g_{zz},\\
g_{m_z}&=-2g_{z\perp}+g_{0z}-2g_{\perp z}+4g_{\perp\perp}-2g_{\perp0}+g_{z0}-2g_{0\perp}+g_{zz}, \\
g_m&=-2g_{z\perp}+g_{0z}-2g_{\perp z}+4g_{\perp\perp}-2g_{\perp0}+g_{z0}-2g_{0\perp}-7g_{zz}
+(m_*/\pi)V_\text{IL}(0).
\end{align}
The inhomogeneous term $\delta_0$  in Eq.~(\ref{gengeq1}) is a bare gap, which is zero in the absence
of external electric field. An analysis of the integral equations for momentum dependent gaps is
presented in Appendix~\ref{append_k_dep_gaps}. Typical solutions of gap equation~(\ref{gengeq1}) for $\delta_0=0$ and different values of coupling constants $g_\delta$ are presented in
Fig.~\ref{fig:gaps_g}(a). The gaps monotonically decrease with momentum $|\mathbf{k}|$ approximately as $|\mathbf{k}|^{-1/2}$.
In the approximation of momentum independent gaps, the integral equations are transformed into
algebraic ones and we can compare the gap sizes of different states without actually solving the corresponding gap equations. Indeed, the solution of the gap equation~(\ref{gengeq1}) for a general
gap parameter $\delta$ is a monotonously increasing function of $g_\delta$, see Fig.~\ref{fig:gaps_g}(b) [the same is true for momentum-dependent $\delta(k)$]. Therefore, the ratio of the gap sizes  in the different states is determined solely by the values of the effective local interaction constants
$g_m$, $g_{m_z}$, $g_{\Delta_z}$, and $g_{\Delta}$. Note that for the QVH phase, the corresponding
local interaction term $g_m$ is effectively reduced because of the negative contribution $(m_*/\pi)V_\text{IL}(0)\equiv\vt\simeq-0.71/\kappa$.

\begin{figure}[h]
\includegraphics[width=55.2mm,valign=t]{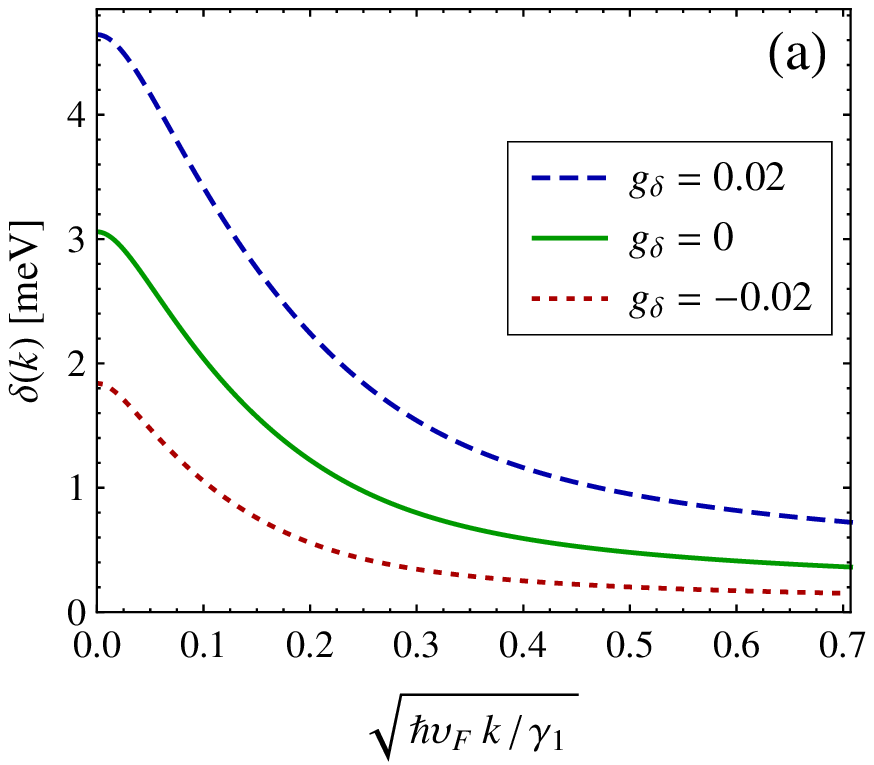}\hspace{1.8mm}
\includegraphics[width=59mm,valign=t]{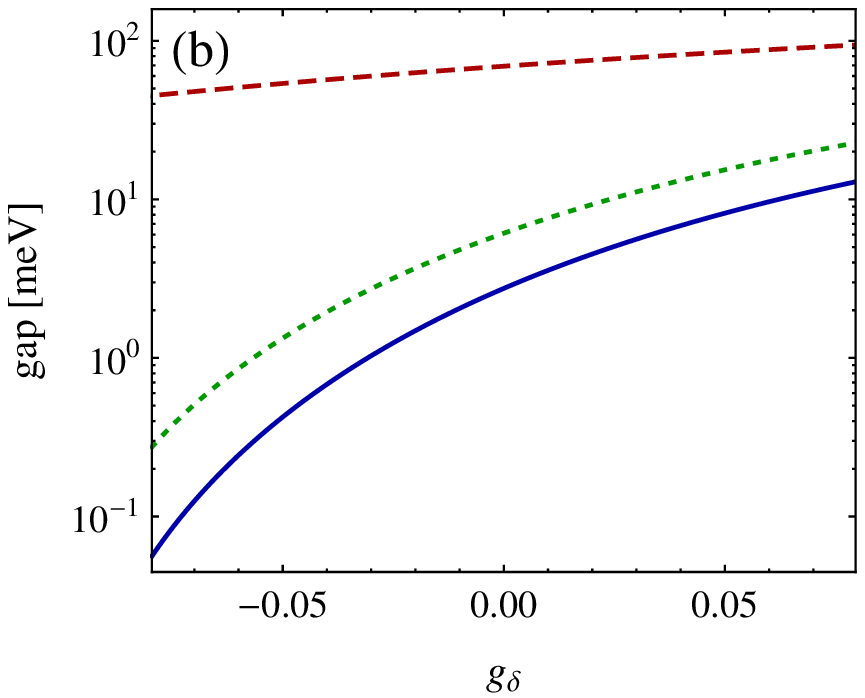}\hspace{4.5mm}
\includegraphics[width=56.5mm,valign=t]{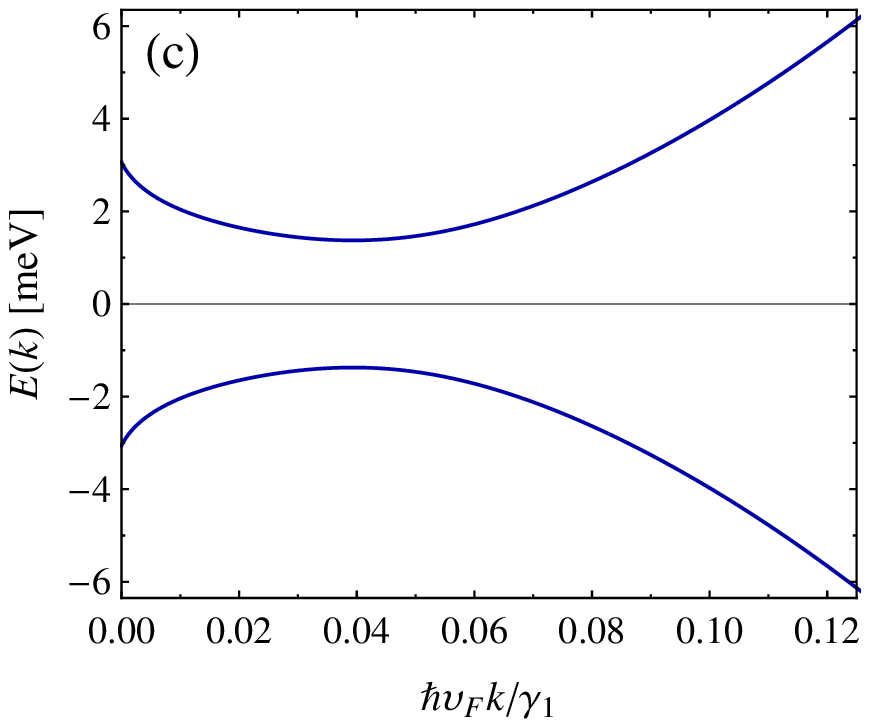}
\caption{(a) Solutions of gap equation~(\ref{gengeq1}) for $\delta_0=0$ and different values of
$g_\delta$. (b) The
dependence of the band gap (solid line) and $2\delta(0)$ (dotted line) on the coupling constant $g_{\delta}$ shown on a logarithmic scale; the dashed line shows the band gap $2\delta$
obtained within the approximation $\delta=\text{const}$. (c) The energy spectrum of the LAF state with $g_{m_z}=0$.
The value $\kappa=1$ is used.}
\label{fig:gaps_g}
\end{figure}

One can see from Fig.~\ref{fig:gaps_g}(b) that, for small symmetry-breaking coupling constants $|g_{\alpha\beta}|\ll1$, the gap magnitude in the absence of external fields is a few meV, which agrees
well with the experimental values of the band gap $0.6$--$3$\,meV~\cite{Weitz,Freitag,Velasco,Bao} (this corresponds to
$-0.04\lesssim g_\delta\lesssim0.003$). The momentum dependence of gaps modifies the dispersion law for quasiparticle excitations,
see Fig.~\ref{fig:gaps_g}(c), the energy of which now has a nontrivial minimum or maximum at nonzero $|\mathbf{k}|$.

In order to determine which of the broken symmetry states realizes the ground state of the system, their energy densities should be calculated by using Eq.~\refer{energy-density-momentum}. The energy densities
for the states with $\delta$ denoting $\Delta_z,\Delta,m$, and $m_z$ are given by
\be
{\cal E}_0=-\int\frac{\dd\omega\dd^2{k}}{2\pi^3}\left(\frac{2E_k^2+\delta^2}
{\omega^2+E_k^2+\delta^2}-\frac{2E_k^2}{\omega^2+E_k^2}\right)=-C\frac{m_*\delta^2(0)}{2\pi},
\label{energy-densities-parallel}
\ee
where the subscript in ${\cal E}_0$ refers to the fact that this is the energy density to the zeroth order
in the perpendicular magnetic field. Further, $C=1$ in the approximation $\delta(p)=\text{const}$, while
the momentum dependence of $\delta(p)$ modifies it to $C\simeq0.17+1.5g_\delta$. Equation~(\ref{energy-densities-parallel}) implies that the state with the largest gap has the lowest energy density and, therefore, is favored. Provided that $g_{\alpha\beta}\ll|\vt|$, the gap in the QVH state
is smaller than gaps of the other states and, thus, the QVH phase is less energetically favorable.
Further, we assume that the linear combination of the local interaction constants corresponding to the
LAF phase is the largest one ($g_{m_z}>g_{\Delta_z},g_{\Delta}$) so that the LAF phase is the most
favorable one. This assumption \cite{Kharitonov-AF} is based on the interpretation of the ground state
in the strong magnetic field as the canted antiferromagnetic state \cite{Kharitonov,Kharitonov1,Maher}
and the experimentally observed continuous evolution of this state when the magnetic field is reduced to zero \cite{Velasco}.

\subsection{Turning on electric field}

When the electric field is turned on, gaps of various states have to mix in order to satisfy the gap equations. In particular, since the matrix structure of the term with $m_0$ in Eq.(\ref{interaction1}) generated by the external electric field in the gap equation is the same as the matrix structure of
the gap of the QVH state, the order parameters of the QSH, QAH, and LAF states mix with the order parameter of the QVH state. The gap equations for these mixed QAH, QSH, and LAF states are
\bea
\delta(p)\pm m(p)&=&\pm m_0+\int\frac{\dd\omega\dd^2{k}}{(2\pi)^3} \frac{\delta(k)\pm m(k)}
{\omega^2+E_k^2+[\delta(k)\pm m(k)]^2}\left[V_{\mbox{\scriptsize eff}}
(\omega,\mathbf{p}-\mathbf{k})+\frac{2\pi(g_\delta+g_m)}{m_*}
\right] \nonumber\\
&+& \int\frac{\dd\omega\dd^2{k}}{(2\pi)^3} \frac{\delta(k)\mp m(k)}
{\omega^2+E_k^2+[\delta(k)\mp m(k)]^2}\frac{2\pi(g_\delta-g_m)}{m_*},
\qquad
\delta=\Delta,\Delta_z,m_z.
\label{gengeq3}
\eea
In addition, the pure QVH state without any mixing still admits solutions even when the
electric field is present. The gap equation for this state is given by
\be
m(p)=m_0+\int\frac{\dd\omega\dd^2{k}}{(2\pi)^3}\frac{m(k)}{\omega^2+E_k^2+m^2(k)}
\left[V_{\mbox{\scriptsize eff}}(\omega,\mathbf{p}-\mathbf{k})+\frac{4\pi}{m_*}g_m\right].
\label{qvhgeq3}
\ee
Note that, as expected, Eq.~\refer{gengeq3} reduces to Eq.~\refer{gengeq1} with $\delta_0=0$ when $m$ and $m_0$ are set to zero.

In order to determine the ground state, we should compare the energy densities of these states, which are determined by
\be
{\cal E}_0=-2\pi m_*\int\frac{d\omega}{2\pi}\int^{\Lambda}\frac{dE_k}{(2\pi)^2}   \left[
\frac{ \Delta_{+}^2+2E_k^2+m_0\Delta_{+}}{\omega^2+\Delta_{+}^2 +E_k^2}
+\frac{ \Delta_{-}^2+2E_k^2+m_0\Delta_{-}}{\omega^2+\Delta_{-}^2 +E_k^2}-\frac{4E_k^2}{\omega^2+E_k^2}\right],
\label{mixed-energy-density}
\ee
where $\Delta_\pm =m\pm\Delta_z$ for the QSH state, $\Delta_\pm =m\pm\Delta$ for the QAH state,
$\Delta_\pm =m\pm m_z$ for the LAF state, and finally $\Delta_\pm =m$ for the QVH state.

Since the gap equations~(\ref{gengeq3}) for different states differ only by the corresponding local
interaction constants, one can easily see that, similarly to the case $E_\perp=0$, the most favorable
among the QSH, QAH, and LAF states is the state with the largest $g_{\delta}$ (which, according to our
assumption, corresponds to the LAF state). On the other hand, a perpendicular electric field $E_\perp$ is expected
to favor the QVH state. When the electric field is weak, the term $m_0$ is only a perturbation to the
gap equations~\refer{gengeq1}. Therefore, the LAF state continues to have lower energy density than
that of the QVH state for a certain range of $m_0$. Whether the QVH state can eventually have lower
energy density as $m_0$ varies to larger values depends on whether this state can lower its energy
density faster than that of the LAF state.

We numerically solved the momentum-dependent gap equations for the LAF and QVH states and calculated
their energy densities as functions of electric field $E_\perp$.
The results are shown in Fig.~\ref{figedo0} for $\kappa=1$. At $E_\perp=0$, our choice of the local four-fermion
coupling constants ensures that the LAF state has the lowest energy density [see Fig.~\ref{figedo0}(a)] and,
therefore, is the ground state. As the electric field becomes larger, the LAF solution ceases to exist
and the pure QVH state becomes the ground state of the system at certain
critical field $E_\perp^\text{cr}$ which depends on the values of $g_{m_z}$ and $g_m$. For the values
of $g_{m_z}$ and $g_m$ in Fig.~\ref{figedo0}(c), the phase transition involves a jump discontinuity in the gaps  [see Fig.~\ref{figedo0}(b)] and thus is the first-order one.
The critical field at $g_{\alpha\beta}=0$ ($g_{m_z}=0$, $g_m=\vt$) is $E_\perp^\text{cr}\simeq7.8$\,mV/nm. If the value of $g_{m_z}$ (which is the only parameter determining
the size of the band gap at $E_\perp=0$) is fixed, then the magnitude of the critical electric field is controlled solely by the coupling constant $g_m$. For example, if $g_{m_z}=0$ (corresponding to the gap $2.7$\,meV), the experimental value $E_\perp^\text{cr}\simeq15-20$\,mV/nm \cite{Weitz,Velasco} implies $-1.1\lesssim g_m-\vt\lesssim-0.66$.

\begin{figure}[ht]
\includegraphics[width=57.5mm,valign=t]{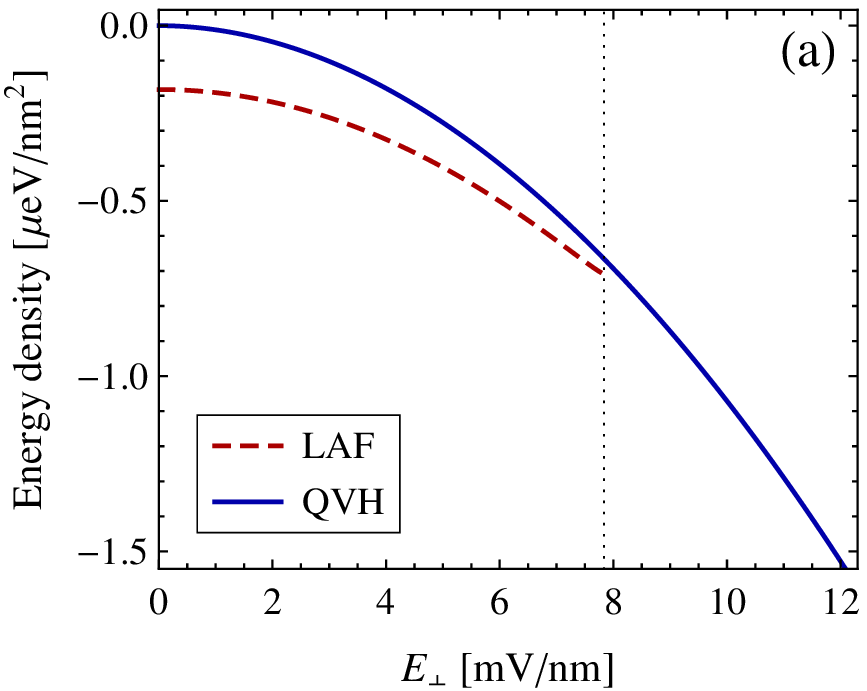}\hspace{3mm}
\includegraphics[width=54.5mm,valign=t]{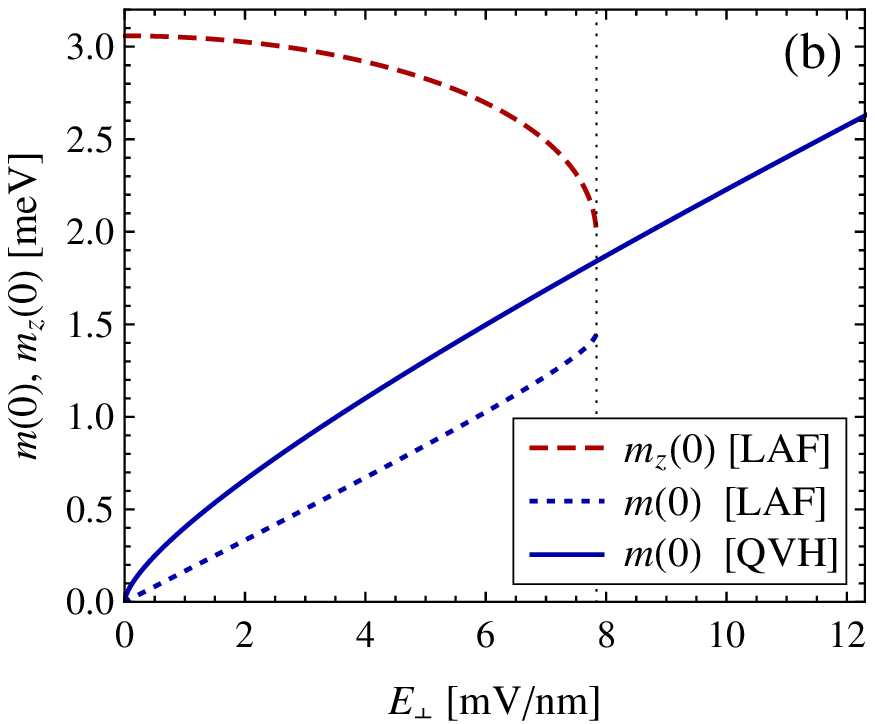}\hspace{4.3mm}
\includegraphics[width=54.8mm,valign=t]{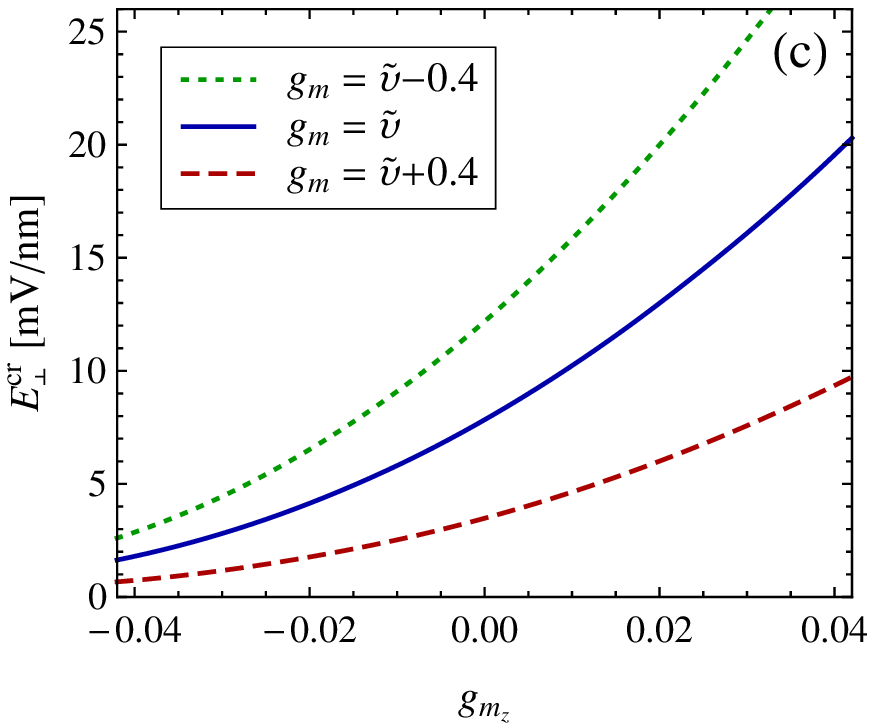}
\caption{The energy densities (a) and the gap parameters at $k=0$ (b) of the LAF and QVH states in
external electric field for $g_{m_z}=0$, $g_m=\vt$.  Panel (c) shows the dependence of the critical field $E_\perp^\text{cr}$ for the LAF-QVH phase transition [vertical dotted line in panels (a) and (b)] on the local interaction constants $g_{m_z}$ and $g_m$. }
\label{figedo0}
\end{figure}

\section{Solutions in in-plane magnetic field}
\label{secb}

In this section, we will study phases of the system in in-plane magnetic field and electric field perpendicular to the planes of graphene. Without the loss of generality, we can assume that the in-plane magnetic field is in the $x$ direction $B_x\neq0$; then the Zeeman interaction $Z\sigma_x$, $Z\equiv B_x\mu_B$, implies that the generalized chemical potential $\mu_x$ should be included in the
analysis for all states.

\subsection{Solutions in the absence of electric field}

Let us analyze solutions of the gap equation first in the case where electric field is absent.
The corresponding ans\"atze for the QVH, LAF, and QSH states can be written as follows:
\bea
&&\tilde{\Sigma}^{\mbox{\scriptsize QVH}}=-\mu_x\sigma_x
-m\eta_3\tau_3\,,\label{qvhbx}\\
&&\tilde{\Sigma}^{\mbox{\scriptsize QAH}}=-\mu_x\sigma_x-\Delta\tau_3\,,\\
&&\tilde{\Sigma}^{\mbox{\scriptsize LAF}}=-\mu_x\sigma_x-\mu_z\sigma_z-(m_x\sigma_{x}+m_z\sigma_{z})\eta_3\tau_3\,,\\
&&\tilde{\Sigma}^{\mbox{\scriptsize QSH}}=-\mu_x\sigma_x-\mu_z\sigma_z
-\left(\Delta_x\sigma_x+\Delta_z\sigma_{z}\right)\tau_3\,. \label{QSH-state-general-ansatz}
\eea
Note that the generalized chemical potentials and gaps of both $x$ and $z$ directions in
spin space are present in the ans\"atze for the LAF and QSH states. In the absence of an in-plane
magnetic field, the Zeeman term vanishes and the direction of gaps in the spin space can be chosen arbitrarily.
For $\mathbf{B}_{||} \ne 0$, the directions along the magnetic field and perpendicular to it are
physically different, therefore, both directions should be present in the most general ans\"atze for
these states.
For the QVH and QAH phases, the gap equations are
\begin{align}
\mu_x(p)-Z&=\frac12\int\frac{d\omega d^2k}{(2\pi)^3}\biggl[V_\text{eff}(\omega,\bp-\bk)+\frac{4\pi}{m_*}g_\mu\biggr]
\sum_{\lambda=\pm}\frac{\lambda\cE_{\lambda}(k)}{\omega^2+\cE^2_{\lambda}(k)},
\\
\delta(p)&=\frac12\int\frac{d\omega d^2k}{(2\pi)^3}\biggl[V_\text{eff}(\omega,\bp-\bk)+\frac{4\pi}{m_*}g_\delta\biggr]
\frac{\delta}{\sqrt{E_k^2+\delta^2}}\sum_{\lambda=\pm}
\frac{\cE_{\lambda}(k)}{\omega^2+\cE^2_{\lambda}(k)}.
\end{align}
with
\begin{equation}
g_\mu=2g_{z\perp}+g_{0z}+2g_{\perp z}+4g_{\perp\perp}+2g_{\perp0}+g_{z0}+2g_{0\perp}+g_{zz},
\end{equation}
and the energy density given by
\begin{equation}
\mathcal E_0=-2\int\frac{d\omega d^2k}{(2\pi)^3}\Biggl\{
\sum_{\lambda=\pm}\biggl[2E_k^2+\delta^2+\mu_x(Z+\mu_x)
+\frac{\mu_x(3E_k^2+2\delta^2)+Z(E_k^2+\delta^2)}{\lambda\sqrt{E_k^2+\delta^2}}\biggr]
\frac{1}{\omega^2+\cE^2_\lambda(k)}-\frac{4E_k^2}{\omega^2+E_k^2}\Biggr\},
\end{equation}
where $\pm\cE_{\pm}(k)$ are four doubly degenerate branches of the energy dispersion, $\cE_{\pm}(k)\equiv\pm\mu_x+\sqrt{E_k^2+\delta^2(k)}$, $\delta(p)=m(p)$ for the QVH phase and $\delta(p)=\Delta(p)$ for the QAH phase.

For the QSH and LAF state, one has the following set of gap equations ($\delta_x=\Delta_x$, $\delta_z=\Delta_z$ for the QSH phase and $\delta_x=m_x$, $\delta_z=m_z$ for the LAF phase)
\bea
\mu_x(p)-Z&=&\frac12\int\frac{\dd\omega\dd^2{k}}{(2\pi)^3}
\left[V_{\mbox{\scriptsize eff}}(\omega,\mathbf{p}-\mathbf{k})+\frac{4\pi}{m_*}g_\mu\right]
\sum_{\lambda=\pm}\frac{\mu_x+\lambda b^{-1}[E_k^2\mu_x+\delta_x(\delta_x\mu_x+\delta_z\mu_z)]}{\omega^2+\cE^2_\lambda(k)},
\label{qshmuxeq}\\
\mu_z(p)&=&\frac12\int\frac{\dd\omega\dd^2{k}}{(2\pi)^3}
\left[V_{\mbox{\scriptsize eff}}(\omega,\mathbf{p}-\mathbf{k})+\frac{4\pi}{m_*}g_\mu\right]
\sum_{\lambda=\pm}\frac{\mu_z+\lambda b^{-1}[E_k^2\mu_z+\delta_z(\delta_x\mu_x+\delta_z\mu_z)]}{\omega^2+\cE^2_\lambda(k)},
\label{qshmuzeq}\\
\delta_z(p)&=&\frac12\int \frac{\dd\omega\dd^2{k}}{(2\pi)^3}
\left[V_{\mbox{\scriptsize eff}}(\omega,\mathbf{p}-\mathbf{k})+\frac{4\pi}{m_*}g_{\delta_z}\right]
\sum_{\lambda=\pm}\frac{\delta_z+\lambda b^{-1}\mu_z(\delta_x\mu_x+\delta_z\mu_z)}{\omega^2+\cE^2_\lambda(k)},
\label{qshdzeq}\\
\delta_x(p)&=&\frac12\int \frac{\dd\omega\dd^2{k}}{(2\pi)^3}
\left[V_{\mbox{\scriptsize eff}}(\omega,\mathbf{p}-\mathbf{k})+\frac{4\pi}{m_*}g_{\delta_z}\right]
\sum_{\lambda=\pm}\frac{\delta_x+\lambda b^{-1}\mu_x(\delta_x\mu_x+\delta_z\mu_z)}{\omega^2+\cE^2_\lambda(k)},
\label{qshdxeq}
\eea
where
\begin{equation}
\cE_\pm(k)=\sqrt{E_k^2+\delta_x^2+\delta_z^2+\mu_x^2+\mu_z^2\pm2b},\qquad
b=\sqrt{E_k^2(\mu_x^2+\mu_z^2)+(\mu_x\delta_x+\mu_z\delta_z)^2},
\end{equation}
with the energy density
\begin{align}
\mathcal E_0&=-2\int\frac{d\omega d^2k}{(2\pi)^3}\Biggl\{
\sum_{\lambda=\pm}\biggl[2E_k^2+\delta_x^2+\delta_z^2+\mu_x(Z+\mu_x)+\mu_z^2
\nonumber \\
&\quad+\frac{E_k^2[Z\mu_x+3(\mu_x^2+\mu_z^2)]
+(\delta_x\mu_x+\delta_z\mu_z)[\delta_x(Z+2\mu_x)+2\delta_z\mu_z]}{\lambda b}\biggr]
\frac{1}{\omega^2+\cE^2_\lambda(k)}-\frac{4E_k^2}{\omega^2+E_k^2}\Biggr\}.
\label{qshbped}
\end{align}

Clearly, the system of equations (\ref{qshmuxeq})--(\ref{qshdxeq}) immediately implies that
$\mu_x\neq0$ if $Z\neq0$. Solutions of this system can be obtained only numerically.
Our analysis shows that this system permits only gapped solutions with either $\delta_z$ or $\delta_x$ nonzero given by
\begin{alignat}{2}
&\text{``noncollinear'' solution:} \qquad &  \mu_x\neq0,\;\; \delta_z\neq0,\;\; \mu_z=0,\;\; \delta_x=0\,,
\label{qshsol1anz}\\
&\text{``collinear'' solution:} &  \mu_x\neq0,\;\; \delta_x\neq0,\;\; \mu_z=0,\;\; \delta_z=0\,,
\label{qshsol2anz}
\end{alignat}
and one gapless ferromagnetic solution with only $\mu_x$ different from zero.
Solutions with both nonzero $\delta_z$ and $\delta_x$ are absent. The presence of nonzero
$\mu_x$ (magnetization) in both solutions (\ref{qshsol1anz}) and (\ref{qshsol2anz}) means the admixture
of the spin-polarized ferromagnetic order, see Eq.~(\ref{order_parameters}). The energy dispersion is given by four doubly degenerate eigenvalues $\pm\cE_\pm(k)$, where $\cE_\pm(k)=\sqrt{(E_k\pm\mu_x)^2+\delta_z^2}$ for noncollinear phases and $\cE_\pm(k)=\pm\mu_x+\sqrt{E_k^2+\delta_z^2}$ for collinear ones.
Evaluating the energy density, we find that the noncollinear solution has lower energy density than that of the collinear solution. The numerical calculations also reveal that, as in the absence of the magnetic field, our assumption $g_{m_z}>g_{\Delta_z}$ implies that the LAF solutions have lower energies than the QSH ones [see Fig.~\ref{fig:only_B_par}(a)]. Likewise, for $g_{m_z}>g_\Delta,g_m$ the energy of the LAF solution is lower that that of the QAH and QVH ones, therefore, the noncollinear (canted) LAF phase remains the ground state for arbitrary $B_\parallel$. In this solution, the spin densities in the two layers have the opposite components in the $yz$ plane, perpendicular to magnetic field and the equal components along the $\mathbf B$ direction \cite{MacDonald}. We also find that the energy density of the gapless ferromagnetic state is always higher than that of the noncollinear LAF and QSH states. In Fig.~\ref{fig:only_B_par}, we illustrate the magnetic field dependence of the free energies and the gaps of the collinear and noncollinear LAF and QSH phases. While the gapped noncollinear solutions exist at arbitrary magnetic fields, the spectrum gap in the collinear solutions closes at some finite $B_\parallel$ value [Fig.~\ref{fig:only_B_par}(b)].

\begin{figure}[ht]
\includegraphics[width=59.9mm,valign=t]{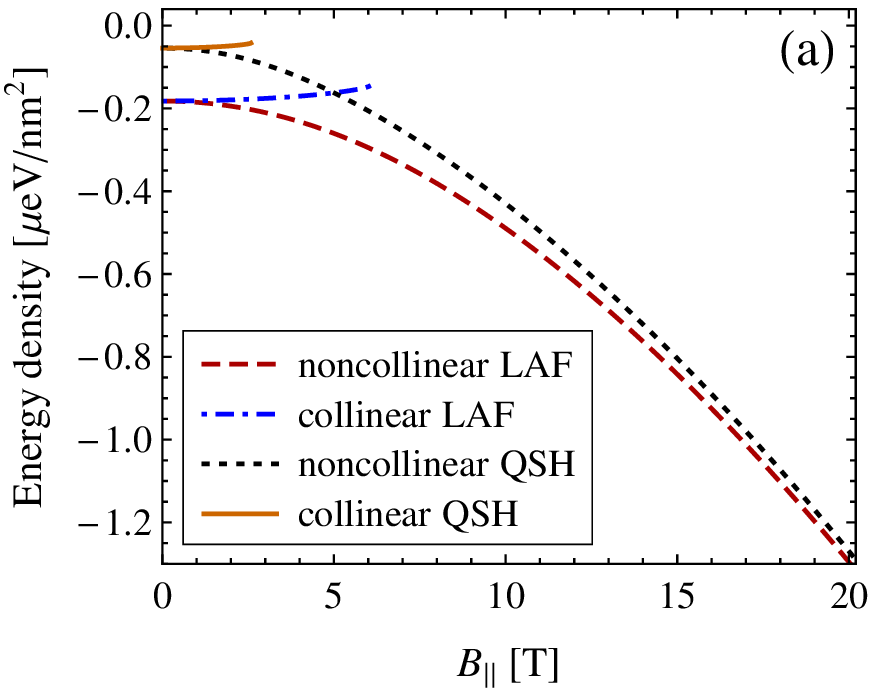}\hspace{3mm}
\includegraphics[width=54.5mm,valign=t]{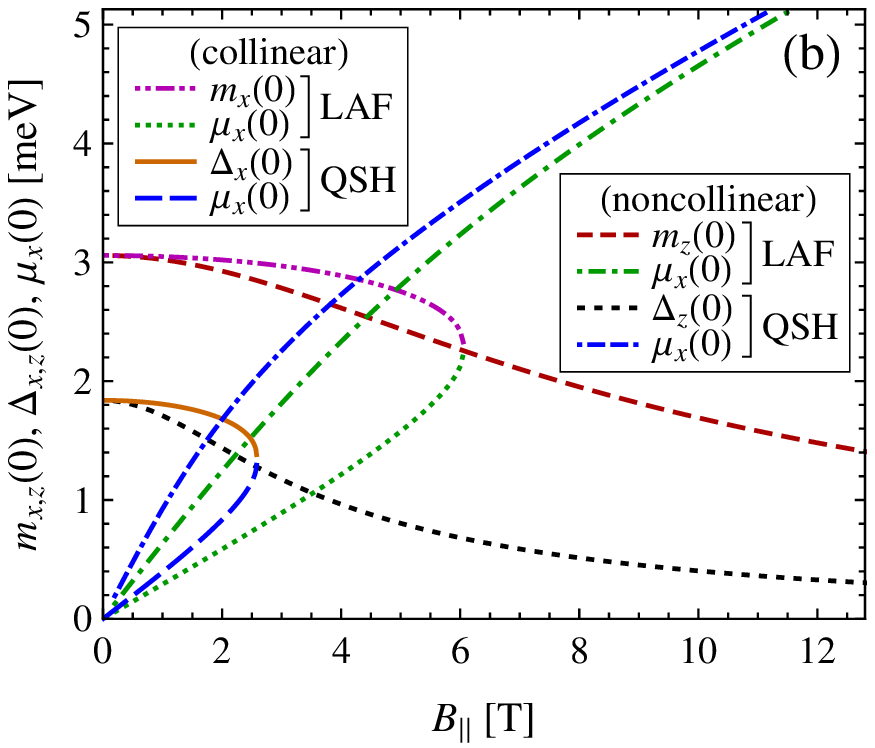}
\caption{The energy densities (a) and gaps at $k=0$ (b) of the LAF and QSH states in
external electric field for $g_{m_z}=0$, $g_{\Delta_z}=-0.02$.}
\label{fig:only_B_par}
\end{figure}

\subsection{Turning on electric field}

In this subsection, we study the phase diagram of the system in electric and in-plane magnetic
fields. We find that it is determined by the competition between the QVH and LAF states.
The corresponding ans\"atze for the QVH, QSH, and LAF states are given by
\bea
&&\tilde{\Sigma}^{\mbox{\scriptsize QVH}}=-\mu_x\sigma_x
-m\eta_{3}\tau_3\,,\\
&&\tilde{\Sigma}^{\mbox{\scriptsize QAH}}=-(\mu_x+\mut_x\eta_3)\sigma_x
-(m\eta_{3}+\Delta)\tau_3\,,\\
&&\tilde{\Sigma}^{\mbox{\scriptsize LAF}}=-\mu_x\sigma_x
-\left(m\eta_{3}+m_z\eta_{3}\sigma_{z}\right)\tau_3\,,\\
&&\tilde{\Sigma}^{\mbox{\scriptsize QSH}}=-\mu_x\sigma_x
-\left(m\eta_{3}+\Delta_z\sigma_{z}\right)\tau_3\,.
\eea
In the above ans\"atze, we consider only the noncollinear LAF and QSH phases, assuming that they are more energetically favorable than their collinear counterparts, similarly to the case $E_\perp=0$. Therefore, compared to Eqs.~(\ref{qvhbx})--(\ref{QSH-state-general-ansatz}) we put $\mu_z=m_x=\Delta_x=0$, and in addition all phases at $E_\perp\ne0$ acquire an additional QVH gap component $m$ (the ans\"atz for the QAH state should also include the generalized chemical potential $\tilde{\mu}_x$ for consistency).

The gap equations for the QVH state have the form
\begin{align}
\mu_x(p)-Z&=\frac12\int\frac{d\omega d^2k}{(2\pi)^3}\biggl[V_\text{eff}(\omega,\bp-\bk)+\frac{4\pi}{m_*}g_\mu\biggr]
\sum_{\lambda=\pm}\frac{\lambda\cE_{\lambda}(k)}{\omega^2+\cE^2_{\lambda}(k)},
\\
m(p)-m_0&=\frac12\int\frac{d\omega d^2k}{(2\pi)^3}\biggl[V_\text{eff}(\omega,\bp-\bk)+\frac{4\pi}{m_*}g_m\biggr]
\frac{m}{\sqrt{E_k^2+m^2}}\sum_{\lambda=\pm}
\frac{\cE_{\lambda}(k)}{\omega^2+\cE^2_{\lambda}(k)},
\end{align}
with the corresponding energy density
\begin{align}
\mathcal E_0&=-2\int\frac{d\omega d^2k}{(2\pi)^3}\Biggl\{
\sum_{\lambda=\pm}\biggl[2E_k^2+m(m_0+m)+\mu_x(Z+\mu_x)
\nonumber \\
&\quad+\frac{\mu_x(3E_k^2+2m^2+mm_0)+Z(E_k^2+m^2)}{\lambda\sqrt{E_k^2+m^2}}\biggr]
\frac{1}{\omega^2+\cE^2_\lambda(k)}-\frac{4E_k^2}{\omega^2+E_k^2}\Biggr\},
\end{align}
where $\pm\cE_\pm(k)$, $\cE_{\pm}(k)\equiv\pm\mu_x+\sqrt{E_k^2+m^2}$, are four doubly degenerate
branches of the energy spectrum.

The gap equations for the QAH state read
\begin{align}
\mu_\pm(p)-Z&=\frac12\int\frac{d\omega d^2k}{(2\pi)^3}\sum_{\lambda,\rho=\pm}
\biggl[\frac{1\pm\rho}{2}V_\text{eff}(\omega,\bp-\bk)+\frac{2\pi}{m_*}(g_\mu\pm\rho g_{\mu_z})\biggr]
\frac{\lambda\cE_{\lambda\rho}(k)}{\omega^2+\cE^2_{\lambda\rho}(k)},
\\
\Delta_\pm(p)\mp m_0&=\frac12\int\frac{d\omega d^2k}{(2\pi)^3}\sum_{\lambda,\rho=\pm}
\biggl[\frac{1\pm\rho}{2}V_\text{eff}(\omega,\bp-\bk)+\frac{2\pi}{m_*}(g_\Delta\pm\rho g_m)\biggr]
\frac{\Delta_\rho}{\sqrt{E_k^2+\Delta_\rho^2}}
\frac{\cE_{\lambda\rho}(k)}{\omega^2+\cE^2_{\lambda\rho}(k)},
\end{align}
and the corresponding energy density is
\begin{align}
\mathcal E_0&=-\int\frac{d\omega d^2k}{(2\pi)^3}\Biggl\{
\sum_{\lambda,\rho=\pm}\biggl[2E_k^2+\Delta_\rho(\Delta_\rho+\rho m_0)
+\mu_\rho(Z+\mu_\rho) \nonumber \\
&\quad+\frac{\mu_\rho(3E_k^2+2\Delta_\rho^2+\rho m_0\Delta_\rho)
+Z(E_k^2+\Delta_\rho^2)}{\lambda\sqrt{E_k^2+\Delta_\rho^2}}\biggr]
\frac{1}{\omega^2+\cE^2_{\lambda\rho}(k)}-\frac{8E_k^2}{\omega^2+E_k^2}\Biggr\},
\end{align}
where $\pm\cE_{\pm\pm}(k)$ are eight energy eigenvalues, $\cE_{\pm,\rho}(k)=\pm\mu_\rho+\sqrt{E_k^2+\Delta_\rho^2}$, $\Delta_\pm\equiv\Delta\pm m$, $\mu_\pm\equiv\mu_x\pm\mut_x$, and
\begin{equation}
g_{\mut_z}=2g_{z\perp}+g_{0z}-2g_{\perp z}-4g_{\perp\perp}-2g_{\perp0}+g_{z0}+2g_{0\perp}+g_{zz}.
\end{equation}
For the LAF and QSH states, the gap equations are
\begin{align}
\mu_x(p)-Z&=\frac12\int\frac{d\omega d^2k}{(2\pi)^3}\biggl[V_\text{eff}(\omega,\bp-\bk)+\frac{4\pi}{m_*}g_\mu\biggr]
\sum_{\lambda=\pm}
\biggl(1+\frac{\lambda(E_k^2+m^2)}{\sqrt{E_k^2\mu_x^2+m^2(\delta_z^2+\mu_x^2)}}\biggr)
\frac{\mu_x}{\omega^2+\cE^2_{\lambda}(k)},
\\
m(p)-m_0&=\frac12\int\frac{d\omega d^2k}{(2\pi)^3}\biggl[V_\text{eff}(\omega,\bp-\bk)+\frac{4\pi}{m_*}g_m\biggr]
\sum_{\lambda=\pm}\biggl(1+\frac{\lambda(\delta_z^2+\mu_x^2)}{\sqrt{E_k^2\mu_x^2+m^2(\delta_z^2+\mu_x^2)}}\biggr)
\frac{m}{\omega^2+\cE^2_{\lambda}(k)},
\\
\delta_z(p)&=\frac12\int\frac{d\omega d^2k}{(2\pi)^3}\biggl[V_\text{eff}(\omega,\bp-\bk)+\frac{4\pi}{m_*}g_{\delta_z}\biggr]
\sum_{\lambda=\pm}\biggl(1+\frac{\lambda m^2}{\sqrt{E_k^2\mu_x^2+m^2(\delta_z^2+\mu_x^2)}}\biggr)
\frac{\delta_z}{\omega^2+\cE^2_{\lambda}(k)},
\end{align}
where $\delta=m_z$ for the LAF state and $\delta=\Delta_z$ for the QSH state. Here $\pm\cE_\pm(k)$,
\begin{equation}
\cE_{\pm}(k)=\sqrt{E_k^2+\delta_z^2+m^2+\mu_x^2\pm2\sqrt{E_k^2\mu_x^2+m^2(\delta_z^2+\mu_x^2)}},
\end{equation}
are four doubly degenerate branches of the energy dispersion.
The energy density is given by
\begin{align}
\mathcal E_0&=-2\int\frac{d\omega d^2k}{(2\pi)^3}\Biggl\{
\sum_{\lambda=\pm}\biggl[2E_k^2+\delta_z^2+m(m_0+m)+\mu_x(Z+\mu_x) \nonumber \\
&\quad+\frac{\mu_x^2(3E_k^2+2m^2+mm_0)+Z\mu_x(E_k^2+m^2)+\delta_z^2m(2m+m_0)}
{\lambda\sqrt{E_k^2\mu_x^2+m^2(\delta_z^2+\mu_x^2)}}\biggr]
\frac{1}{\omega^2+\cE^2_\lambda(k)}-\frac{4E_k^2}{\omega^2+E_k^2}\Biggr\}.
\end{align}
Solutions of these gap equations can be obtained only numerically. In general, we found that all
gaps weakly depend on $B_\parallel$ and the main dependence on $B_\parallel$ is contained in $\mu_x$,
see Fig.~\ref{fig:EcritB}(c). The generalized chemical potential $\mu_x$ for the QSH and LAF states
has an almost linear dependence on the magnetic field.

\begin{figure}[h]
\includegraphics[width=57.5mm,valign=t]{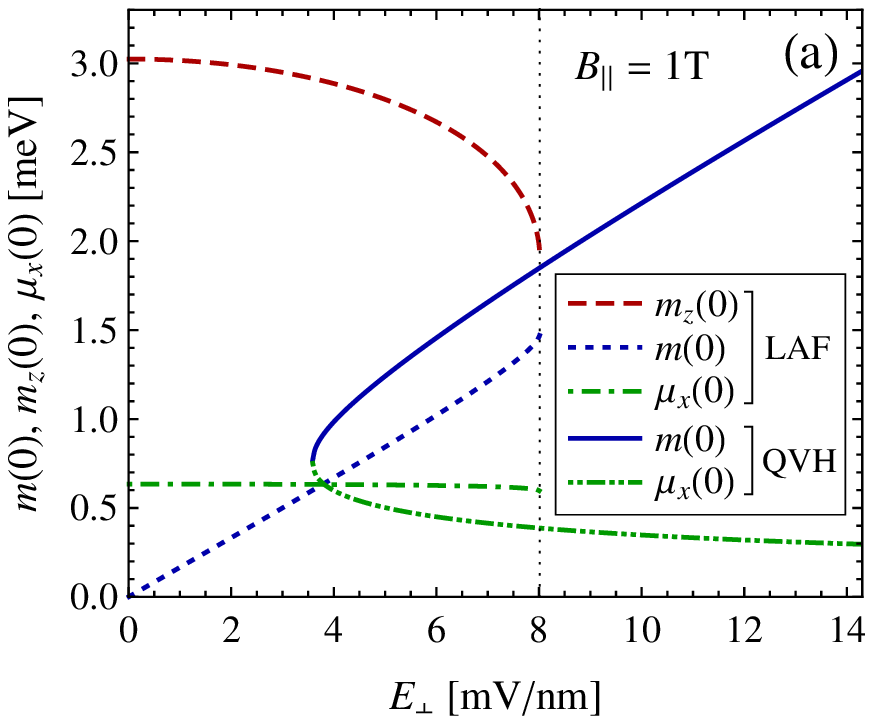}\hspace{3.7mm}
\includegraphics[width=55.3mm,valign=t]{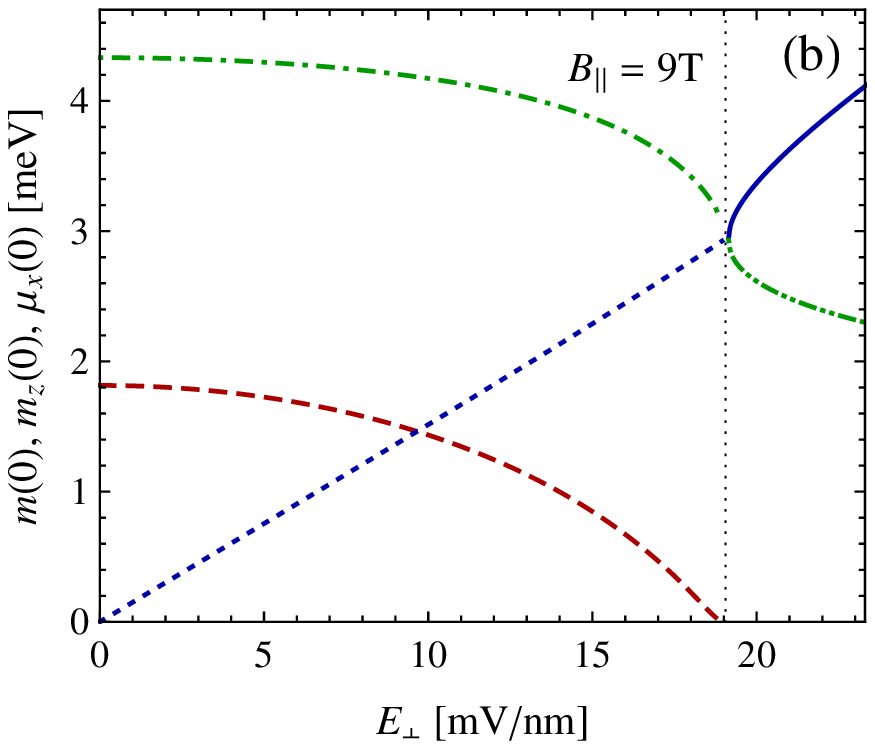}
\bigskip \\
~\hspace{1mm}\includegraphics[width=55.3mm,valign=t]{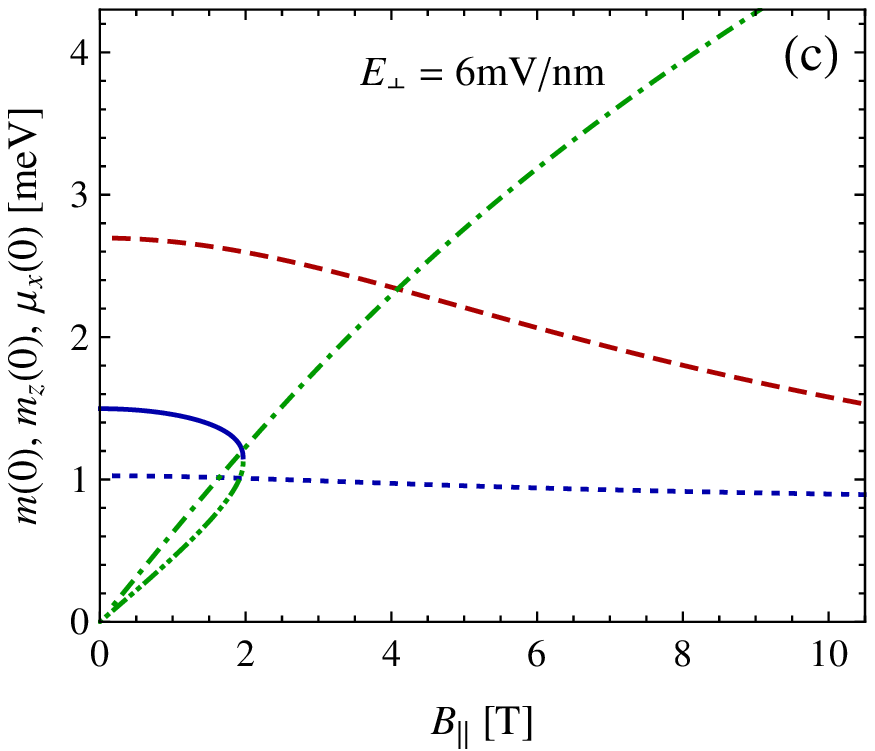}\hspace{2.6mm}
\includegraphics[width=56.5mm,valign=t]{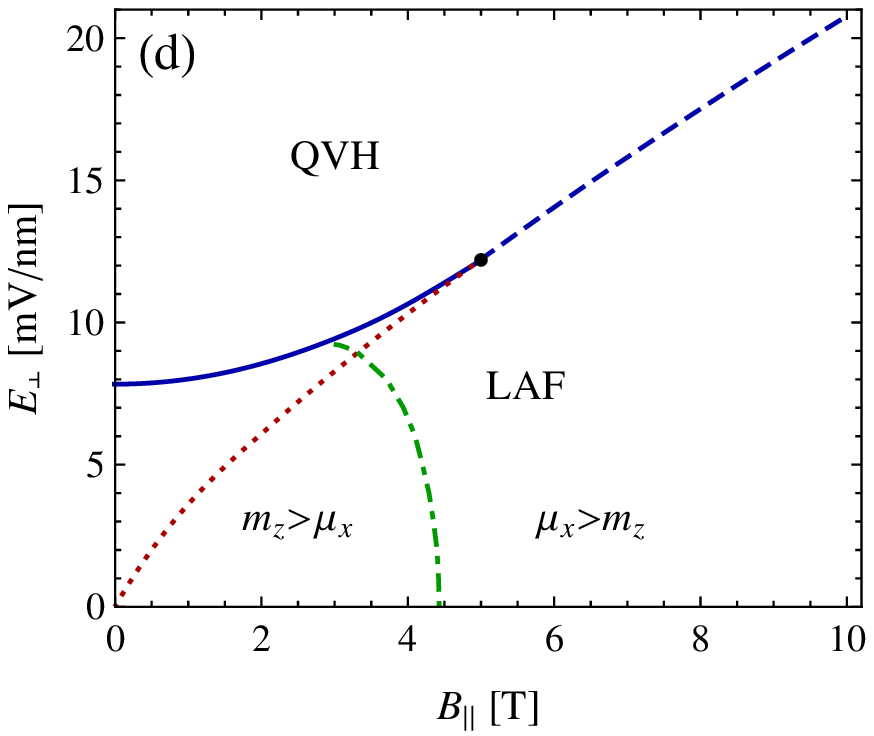}
\caption{The gaps at $k=0$ of the LAF and QVH states as functions of
external electric field for $B_\parallel=1$\,T~(a) and $B_\parallel=9$\,T~(b), of parallel magnetic field for $E_\perp=6$\,mV/nm (c) and the phase diagram in electric and in-plane magnetic fields (d). Parameters used: $g_{m_z}=g_{\mu}=0$, $g_m=\vt$, and $\kappa=1$. The line types used in panels (a)--(c) are described in the inset of panel (a).
In panel~(d), the solid and dashed lines describe the first and second order phase transitions between
the LAF and QVH states. The black point marks the critical point where the line of the first order phase transition terminates. The dotted line shows the boundary of the region of existence of the gapped
QVH state and the dot-dashed line splits the LAF into regions where $m_z(k)$ or $\mu_x(k)$ dominates.}
\label{fig:EcritB}
\end{figure}

Having found the generalized chemical potentials and gaps, we can calculate and compare the energy
densities of various states and then determine the phase diagram of the system. We find that
the external electric and parallel magnetic fields do not change the ordering of the energies of
the QAH, QSH, and LAF solutions and thus the ground state is not QAH or QSH. In Figs.~\ref{fig:EcritB}(a) and~\ref{fig:EcritB}(b) we plot the zero-momentum gaps for  the LAF and QVH states as functions
of an electric field for two values of the in-plane magnetic field, $B_\parallel=1$\,T and $B_\parallel=9$\,T, respectively. The largest gaps determine the system ground states.
The corresponding phase diagram of bilayer graphene in electric and in-plane magnetic fields for the zero-momentum gaps is plotted in Fig.~\ref{fig:EcritB}(c). The phase boundary, composed of the solid and dashed lines, separates
the pure QVH and mixed LAF states. For the solid line part of the phase boundary ($B_\parallel\lesssim5$\,T),
it is found that the phase transition is of the first order due to gaps changing discontinuously
across the phase transition line while it is of the second order for the dashed line part ($B_\parallel\gtrsim5$\,T). The phase
boundary for $B_\parallel\lesssim5$\,T is determined by a similar mechanism as in the case of zero $B_\parallel$. Again the mixed LAF
phase exists only when external voltage is smaller than a certain critical value $E_\perp^\text{cr}$.
As $B_\parallel$ increases, the critical $E_\perp$ increases also, while the energy density
of the QVH state is not affected much by $B_\parallel$.  The critical line is perfectly fitted by
the quadratic dependence at small $B_\parallel$
\be
E_\perp^\text{cr}=a+b B^2_\parallel,
\label{critical-line}
\ee
where $a\simeq7.8$\,mV/nm and $b\simeq0.18$\,mV/(nm$\times\mathrm{T}^2$) for $g_{\alpha\beta}=0$ and
$\kappa=1$. On the other hand, if the value of $g_{m_z}$ is determined from the experimental value of
the band gap in the absence of external fields and $g_m$ is determined from the experimental value
of $E_\perp^\text{cr}$ at $\mathbf B=0$, then the phase diagram in Fig.~\ref{fig:EcritB} depends on
a single free parameter $g_\mu$. We found that although the values of $a$ and $b$ might change as
$g_{m_z}$, $g_m$, and $g_\mu$ vary, the quadratic dependence of the critical line on the strength of
in-plane magnetic field is preserved.
For $B_\parallel\gtrsim5$\,T, the phase transition becomes a continuous one with the antiferromagnetic order parameter vanishing at the critical line. The latter has approximately linear dependence on $B_\parallel$.

We find that the noncollinear LAF phase is stable with respect to increasing the parallel magnetic field: although for $B_\parallel$ exceeding 3--5\,T the Zeeman-like parameter $\mu_x(k)$ becomes larger that the antiferromagnetic gap $m_z(k)$ [see Figs.~\ref{fig:EcritB}(c) and~\ref{fig:EcritB}(d)], the latter does not vanish and this phase remains the ground state for all experimentally accessible magnetic fields. This behavior
is in consonance with studies in monolayer \cite{Roy} and bilayer \cite{Roy2} graphene where a transition from the easy-plane
antiferromagnet to a pure ferromagnetic phase is not found. For earlier studies of the role
of in-plane magnetic field in specific $(2+1)$-dimensional Gross-Neveu model, see Ref.~\cite{Klimenko}. Experimentally, the gapped ground state in suspended bilayer was found to be stable in parallel magnetic fields at least up to $3$~T in Ref.~\cite{Freitag2}.

The critical line in Fig.~\ref{fig:EcritB} qualitatively agrees with the findings in Ref.~[\onlinecite{Maher}], where the phase diagram of the $\nu=0$ state in bilayer graphene was experimentally studied as a function of perpendicular electric and total magnetic
$B_{\rm{tot}}$ fields with the out-of-plane magnetic field fixed at $B_{\perp}=1.75$\,T. Experimentally, at low $B_{\rm{tot}}$, the
phase boundary between the QVH and CAF phases is practically flat. The CAF state at $B_\perp\ne0$} continuously interpolates between the LAF and ferromagnetic states as the in-plane magnetic field
increases~\cite{Kharitonov,Kharitonov1}, and at small $B_{\perp}$ it is not much different from the
(noncollinear) LAF state considered in our analysis. For $B_{\rm{tot}}$ larger than approximately 15\,T,
the critical electric field separating the QVH and ferromagnetic phases increases with
$B_{\rm{tot}}$ linearly \cite{Maher}. In our analysis, the critical line between the QVH and LAF states for $B_\parallel\gtrsim5$\,T also has a linear shape with approximately the same slope of $1.7$\,mV/(nm$\times$T). Thus, the phase diagram obtained in Ref.~[\onlinecite{Maher}] (see Fig.~4 therein) has many similar features to our phase diagram at $B_\perp=0$. The main difference from Ref.~[\onlinecite{Maher}] is that according to our phase diagram the canted LAF state remains a stable ground state in the absence of out-of-plane magnetic field at large $B_\parallel$: There is no phase transition to a ferromagnetic state with conductance $4e^2/h$ due to topologically protected edge states. Since the LAF state does not have edge states, the absence of a phase transition at $B_\perp=0$ can be checked experimentally.

\section{Linear and quadratic in $B_\perp$ corrections}
\label{5}

In this section, we focus our analysis on the gap generation in a weak perpendicular magnetic field and set, for simplicity, $B_{||}=0$.  We treat $B_\perp$ as a perturbation and consider the corrections to the generalized chemical potentials,
gaps, and energy densities in the linear and quadratic orders in $B_\perp $. The
corresponding results for the fermion propagator are given in Appendix~\ref{gap-eq-in-Bperp}. This perturbative analysis is expected to be valid
when the cyclotron energy $\hbar\omega_c=e\hbar B_\perp/(m_*c)\approx2.2B_\perp$[T]\,meV is much less than the dynamically generated gap at $B_\perp=0$.
Taking into account the experimentally observed band gaps up to $3$\,meV \cite{Weitz,Freitag,Velasco,Bao}, we have the condition $B_\perp\ll1$\,T. In order to simplify our analysis, we will not consider in-plane magnetic field in this section. Moreover, due to the complexity of the gap equation, we will consider
only momentum-independent generalized chemical potentials and gaps. Since this approximation leads to
the systematic overestimation of both the gap magnitudes and $E_\perp^\text{cr}$, we use $\kappa=3$ in
order to keep them closer to the experimental values.

In order to use Eq.~\refer{SD-equation-momentum} and obtain a self-consistent and recursive system of equations, we should first specify the ans\"atz for the QVH and LAF states. Due to the Zeeman term, we
should include $\mu_{03}\equiv\mu_z\neq0$ in ans\"atz~\refer{ansatz} for both the QVH and LAF states in order that the gap equations have consistent solutions. Therefore, the full ans\"atz for these
two states is given by
\begin{alignat}{2}
&\mbox{QVH:}\quad&
\tilde\Sigma=&-m\eta_3\tau_3-B_\perp(\mu^{(1)}_z\sigma_3+\mut^{(1)}\eta_3)-B_\perp^2(m^{(2)}
\eta_3\tau_3+\Delta_z^{(2)}\sigma_3\tau_3),
\label{ansatz-QVH}
\\
&\mbox{LAF:}&
\tilde\Sigma=&-m_x\sigma_1\eta_3\tau_3-m\eta_3\tau_3-B_\perp(\mu^{(1)}_z\sigma_3+\mut^{(1)}_x
\sigma_1\eta_3+\mut^{(1)}\eta_3) \nonumber
\\
&&&-B_\perp^2(m_x^{(2)}\sigma_1\eta_3\tau_3+m^{(2)}\eta_3\tau_3+\Delta_z^{(2)}\sigma_3\tau_3).
\label{ansatz-LAF}
\end{alignat}
Note that the $\mathcal O(B_\perp^0)$ antiferromagnetic gap parameter $m_x$ is perpendicular to the
external magnetic field (in general, it points in an arbitrary direction in the graphene sheet plane),
while the $\mathcal O(B_\perp^1)$ parameter $\mu_z^{(1)}$ leads to a small tilting in the $\mathbf B$ direction. Similarly to the previously considered case of a parallel magnetic field, our numerical calculations show that this noncollinear orientation lowers the energy of the system and is thus favorable.
Hence, the perpendicular magnetic field transforms the purely antiferromagnetic state into the canted antiferromagnetic one \cite{Kharitonov,Kharitonov1}, therefore, in what follows we will use
the  notation CAF for this state. The corresponding gap equations in the quadratic
order in $B_\perp$ for the parameters entering ans\"atze (\ref{ansatz-QVH}), (\ref{ansatz-LAF}) can be obtained from Eq.~\refer{SD-equation-momentum} and are written down in Appendix~\ref{gap-eq-in-Bperp}.

Solving the gap equations numerically and then substituting the solutions in the energy density~\refer{edb2}, we found that the first order in $B_{\perp}$ correction to the energy density
is zero for both the QVH and CAF states and a nonzero contribution is connected with the second order in $B_{\perp}$ correction as
shown in Fig. \ref{figedo2}(b). It is seen that this correction is positive
for the the QVH phase and negative for the LAF phase. Therefore, the point where the energies
cross at $\mathbf B=0$ [see Fig.~\ref{figedo2}(a)] shifts towards the larger $E_\perp$ values with
growing $B_\perp$ while remaining the first order transition. The resulting critical line has the form $E_\perp^\text{cr}=a+b B^2_\perp$ and is plotted in Fig.~\ref{figedo2}(c). Experimentally,
the existence of this phase transition has been proven but the exact expression for this critical line
is still not clear.

\begin{figure}[h]
\includegraphics[width=55.5mm,valign=t]{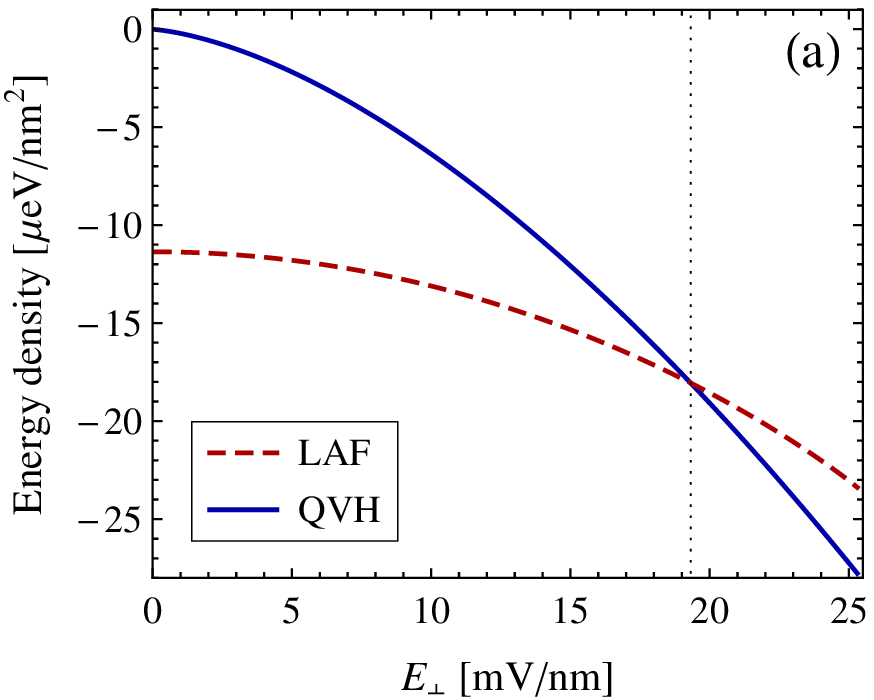}\hspace{3.7mm}
\includegraphics[width=56.1mm,valign=t]{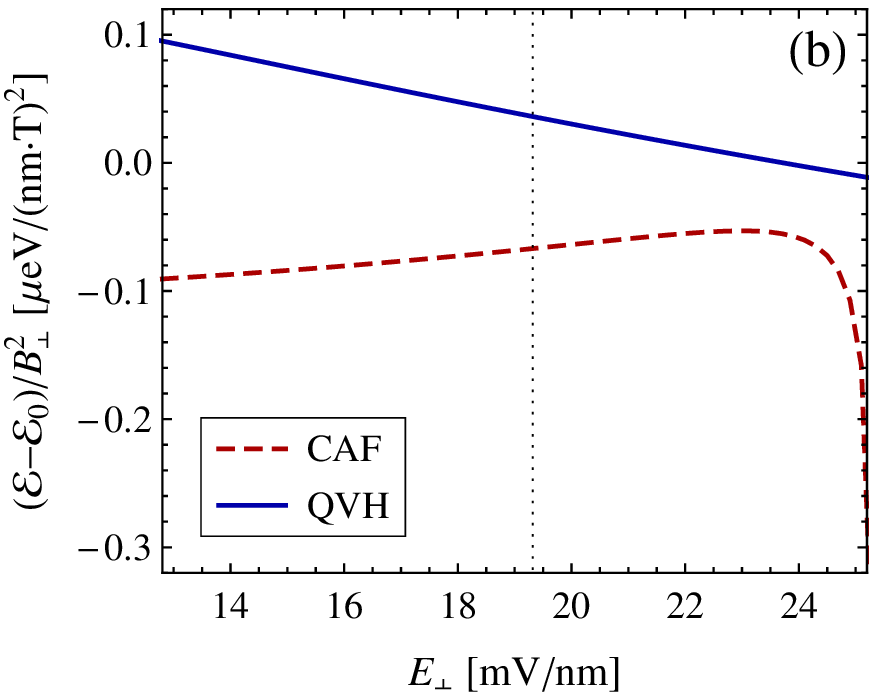}\hspace{5mm}
\includegraphics[width=55.3mm,valign=t]{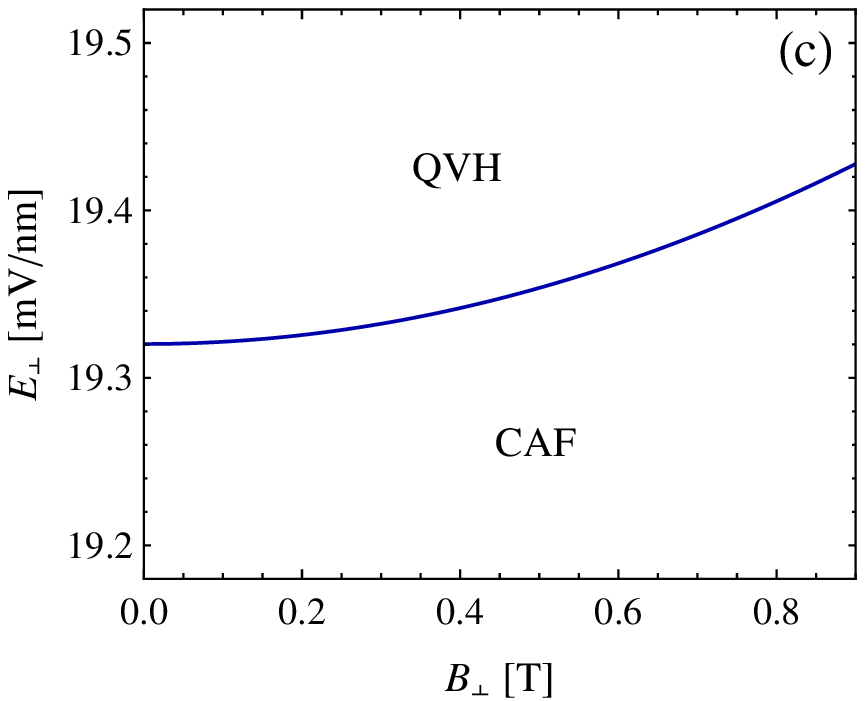}
\caption{The energy densities of the LAF and QVH states at $\mathbf B=0$ (a), their corrections due to finite $B_\perp$ in the vicinity of the phase transition at $E_\perp=E_\perp^\text{cr}$ (marked by the vertical dotted line) (b), and the phase diagram in the $(B_\perp,E_\perp)$ plane (c) for $g_{\alpha\beta}=0$ and $\kappa=3$.}
\label{figedo2}
\end{figure}

\section{Conclusion}
\label{6}

In this paper, we studied the gap generation and dynamical SU(4) symmetry breaking in bilayer
graphene at the neutrality point in electric and in-plane magnetic fields as well as a weak out-of-plane magnetic field $\mathbf{B}_{\perp}$. We utilize a model with the dynamically screened
Coulomb interaction and contact four-fermion interactions which explicitly break the SU(4) valley-spin symmetry. We emphasize the following new points in our analysis. Although it was found \cite{Kharitonov,Kharitonov1,Kharitonov-AF} that the contact four-fermion interactions are crucial for
the correct choice of the ground state of bilayer graphene, a study of broken symmetry states in a model with both the long-range Coulomb interaction and SU(4) asymmetric contact interactions was never done in the literature. Further, beginning with the pioneer work \cite{Levitov}, the approximation of momentum-independent gaps was applied in all studies of the gap equations. This is indeed a reasonable first approximation because the polarization effects in bilayer graphene are much stronger than those in
monolayer graphene and the static screened Coulomb potential in bilayer graphene is approximately constant
in a large interval of momenta. However, it is very important to take into account the frequency dependence of polarization effects which significantly decrease screening. In this case,
our study shows that the momentum dependence of gaps, which can be approximated by $|\mathbf{k}|^{-1/2}$ at large momenta,
is essential and diminishes by an order of magnitude the gaps compared to the case of the momentum-independent approximation.

By numerically solving the momentum-dependent gap equations for the broken symmetry states and determining their energy densities as functions of electric field $E_{\perp}$ and in-plane magnetic field $B_{||}$, we found that  the LAF phase is realized at small values of $E_{\perp}$, while the QVH phase is the ground state of the system at large $E_{\perp}$.
As in-plane magnetic field $B_{||}$ increases, the critical electric field $E^{\rm cr}_{\perp}$ increases too. The part of the critical line separating the LAF and QVH phases at magnetic field $B_{||}\lesssim 5$T  has a quadratic dependence $E^{\rm cr}_{\perp}=a+bB^2_{||}$, and we found that the phase transition across this line
is of the first order due to gaps changing discontinuously. For $B_{||}\gtrsim 5$T, the critical line has
approximately linear dependence on $B_\parallel$ and the phase transition becomes a continuous one.

We show that although the SU(4)-asymmetric contact interactions include, in general, eight independent constants, the gap sizes and energy density of the ground LAF state, experimentally observed in the absence of external fields, are controlled by a single linear combination $g_{m_z}$. Furthermore,
in order to describe other broken symmetry ground states in perpendicular electric and in-plane magnetic fields, one should take into account two additional linear combinations $g_m$ and $g_\mu$. Thus the phase diagram in the $(B_\parallel,E_\perp)$ plane is parametrized by the three independent effective local interaction constants. We found that for some reasonable choice of these parameters the band gap and the critical electric field at $\mathbf B=0$ agree well with the corresponding experimental data available in the literature.
According to our phase diagram, the (canted) LAF  state remains a stable ground state in the absence of out-of-plane magnetic field at large $B_\parallel$: There is no phase transition at low displacement field $E_\perp$. This result is similar to studies in monolayer \cite{Roy} and bilayer \cite{Roy2} graphene where a transition from the antiferromagnetic to a pure ferromagnetic phase was not found.
Since the LAF state does not have edge states, the absence of a phase transition at $B_\perp=0$ driven by in-plane magnetic field can be checked experimentally. Given that existing experiments at $B_\perp\ge1.75$\,T \cite{Maher,Pezzini} support the scenario of the CAF to ferromagnetic state phase transition in the strong parallel fields \cite{Kharitonov1},
it would be interesting to investigate theoretically the fate of this transition in the $(B_\perp,B_\parallel)$ plane at smaller $B_\perp$ values. This question will be addressed elsewhere.

We studied the role of weak perpendicular magnetic field in the particular case of zero in-plane magnetic field.
By using a perturbation theory in a perpendicular magnetic field $B_{\perp}\ll1$T with $B_\parallel=0$, we found that the broken symmetry states and phase diagram are stable and remain qualitatively unchanged. The main consequence of the presence of $B_{\perp}$ is that the LAF state transforms into the CAF state in agreement with previous theoretical and experimental studies. 

Finally, we would like to add that studied broken states may find practical applications.
For instance, recently it was proposed
\cite{Guinea} that the CAF state could be instrumental for the creation of topological superconductivity in graphene-superconductor
junctions without the need for strong spin-orbit coupling. The key advantage of the CAF state is its magnetic ordering due to the contact
four-fermion interactions. Therefore, coupling this state to a conventional superconductor gives rise to Majorana bound states and makes the
CAF state a promising platform for Majorana physics in graphene systems.

\begin{acknowledgments}
We thank V.A. Miransky for fruitful discussions.
The work of Junji Jia is supported by the
Chinese SRFDP 20130141120079, NNSF China 11504276 \& 11547310,
Ministry of Science and Technology of China (Grant No. 2014GB109004),
and Natural Science Foundation of Hubei Province (Grant No. ZRY2014000988).
V.P.G.  acknowledges the support of the RISE Project CoExAN GA644076. The work of E.V.G and V.P.G. was supported partially by the Program of Fundamental Research of the
Physics and Astronomy Division of the NAS of Ukraine.

\end{acknowledgments}

\appendix

\section{Momentum-dependent gap parameters}
\label{append_k_dep_gaps}

In this section we solve the gap equations keeping the momentum dependence of the gap parameters
(the frequency dependence is neglected). Let us start with the gap equation
\begin{equation}
\delta(p)=\delta_0+\int\frac{d\omega d^2k}{(2\pi)^3}\frac{\delta(k)}{\omega^2+E_k^2+\delta^2(k)}
\left[V_\text{eff}(\omega,\bp-\bk)+\frac{4\pi}{m_*}g_\delta\right]
\label{gap_eqn_orig}
\end{equation}
for a general gap parameter $\delta(p)$, where $\delta_0$ is a bare gap and $g_\delta$ is the
corresponding local interaction constant. Using Eqs.~(\ref{veff}) and~(\ref{polarization}), the
above equation can be written as
\begin{equation}
\delta(p)=\delta_0+\frac{1}{4\pi^2}\int\limits_0^\infty d\omega\int\limits_0^{2\pi}d\theta_k
\int\limits_0^\Lambda dE_k\frac{\delta(k)}{\omega^2+E_k^2+\delta^2(k)}
\biggl(\frac{1}{\alpha^{-1}\sqrt{E_{\bp-\bk}/\gamma_1}+2P(\omega/E_{\bp-\bk})}+4g_\delta\biggr),
\label{gap_eqn_2}
\end{equation}
where $\theta_k$ is the angle between vectors $\bk$ and $\bp$, and $\alpha=e^2/(\kappa v_F)
\simeq2.7/\kappa$ is the graphene's effective ``fine structure constant''. The frequency integration
can be done for the approximate form of the polarization function \cite{Levitov}
\begin{equation}
P(z)\approx\frac12\biggl(\frac{z^2}{\pi^2}+\frac1{4\ln^24}\biggr)^{-1/2},
\label{pol_Nandkishore}
\end{equation}
which respects the asymptotics $P(0)=\ln4$, $P(z\gg1)\simeq\pi/(2z)$ of the original polarization function~(\ref{polarization}). Using expression~(\ref{pol_Nandkishore}) for the polarization function
 and the formula
\begin{align}
\int\limits_0^\infty\frac{dx}{(x^2+1)(u+1/\sqrt{1+v^2x^2})}
&=\frac1{u^2(v^2-1)+1}\biggl[\frac{v\arctan\sqrt{u^2-1}}{\sqrt{u^2-1}}+\frac\pi2u(v^2-1)
+\sqrt{1-v^2}\arctan\frac{\sqrt{1-v^2}}{v}\biggr] \\
&\equiv F(u,v),
\end{align}
one gets from Eq.~(\ref{gap_eqn_2})
\begin{equation}
\delta(p)=\delta_0+\frac{1}{2\pi}\int\limits_0^{2\pi}d\theta_k
\int\limits_0^{\Lambda}dE_k\Bigl[K\Bigl(|\bp-\bk|,\sqrt{E_k^2+\delta^2(k)}\Bigr)+g_\delta\Bigr]
\frac{\delta(k)}{\sqrt{E_k^2+\delta^2(k)}},
\end{equation}
where
\begin{equation}
K\bigl(q,f(k)\bigr)=
\frac{1}{4\pi\ln4}F\biggl(\frac{\sqrt{E_q/\gamma_1}}{2\alpha\ln4},
\frac{2\ln4}{\pi E_q}f(k)\biggr).
\end{equation}
The angular integration can be performed if one uses the following approximation for the kernel:
\begin{equation}
K\bigl(|\bp-\bk|,f(k)\bigr)\approx
K\bigl(\max(p,k),f(k)\bigr)\equiv \Kt\bigl(p,k,f(k)\bigr),
\end{equation}
and we finally arrive at
\begin{equation}
\delta(p)=\delta_0+\int\limits_0^{\Lambda}dE_k\Bigl[\Kt\Bigl(p,k,\sqrt{E_k^2+\delta^2(k)}\Bigr)
+g_\delta\Bigr]\frac{\delta(k)}{\sqrt{E_k^2+\delta^2(k)}}.
\label{gen_gap_eqn}
\end{equation}
The above gap equation can be used for the LAF, QSH, QAH, and QVH states in the simplest cases;
for example, with $\delta(p)=m(p)$, $\delta_0=m_0$ it coincides with Eq.~(\ref{qvhgeq3})
for the QVH state at $\mathbf B=0$. The generalization to the case of a system of a few gap equations
is also straightforward. We solve the resulting integral equations of the form~(\ref{gen_gap_eqn}) iteratively by using a discrete momentum grid $k_i$ with a few hundred points, uniform in $k^{1/4}$. Starting from some initial guess for $\delta(k_i)$, we then either evaluate the right-hand side
directly (with the integral calculated by the trapezoidal rule) or solve the resulting system of
linear equations for $\delta(k_i)$ [$\delta(k_i)$ in the denominator of the integrand and in the
kernel $\Kt$ are taken from the previous iteration step], whichever leads to the convergent iterations.

Figure~\ref{fig:gaps_g}(a) shows the momentum-dependent gap $m_z(p)$ in the LAF state with $g_{m_z}=0$
in the absence of external fields, obtained by numerically solving Eq.~(\ref{gen_gap_eqn}) with $g_\delta=0$, $\delta_0=0$. The energy spectrum has the ``mexican-hat'' form, see
Fig.~\ref{fig:gaps_g}(c).

\section{Gap equations in the second order in $B_\perp$}
\label{gap-eq-in-Bperp}

As mentioned in the Introduction, our aim is to study the gap generation in bilayer graphene in
perturbation theory in perpendicular magnetic field with $B_{||}=0$. By making use 
of Eq.~(\ref{self-energy}), let us express the fermion propagator
through the self-energy in the perturbation theory in $B_\perp$. The free
inverse fermion propagator in an external magnetic field is given by
\begin{equation}
S^{-1}(x,z)=\left(\,i\hbar\partial_{x_0}+\frac{\hbar^2}{2m_*}
\left( \begin{array}{cc} 0 &
(-i D_{x_1}-D_{x_2})^2 \\
(-iD_{x_1}+D_{x_2})^2 & 0 \end{array} \right)\,\right)\,\delta^3(x-z)
\end{equation}
with the covariant derivative $D_{k}=\partial_{k}+(ie/\hbar c)A_{k}$ ($e>0$). Further, by writing
\begin{equation}
\int d^3z\,S^{-1}(x,z)G(z,y)=e^{i\Phi(x,y)}
\biggl(i\hbar\partial_{x_0-y_0}+\frac{\hbar^2}{2m_*}
\begin{bmatrix}
0 & (-iD_{x_1-y_1}-D_{x_2-y_2})^2 \\
(-iD_{x_1-y_1}+D_{x_2-y_2})^2 & 0
\end{bmatrix}
\biggr)\tilde G(x-y),
\end{equation}
and using the identity
\begin{equation}
\mathbf{x}\mathbf{A(z)}+\mathbf{z}\mathbf{A(y)}=\mathbf{x}\mathbf{A(y)}+(\mathbf{x}-\mathbf{y})
\mathbf{A(z-y)}\,,
\end{equation}
for the quantity $\mathbf{x}\mathbf{A}(\mathbf{y})=-(B_{\perp}/2)(\mathbf{x}\times\mathbf{y})
\equiv -(B_{\perp}/2)\epsilon_{ij}x_iy_j,\, i,j=1,2$, Eq.~(\ref{self-energy}) in momentum space takes the form
\begin{equation}
\left(\,\omega+ \frac{\hbar^2}{2m_*}\left( \begin{array}{cc} 0 &D^- \\
D^+ & 0 \end{array} \right)\,\right)\tilde{G}(\omega,\mathbf{p})
+\int d^2(\mathbf{x}-\mathbf{y}) e^{-i\mathbf{p}(\mathbf{x}-\mathbf{y})}
\int d^2\mathbf{z}\,e^{-i(e/\hbar c)(\mathbf{x}-\mathbf{y})\mathbf{A(z-y)}}
\tilde{\Sigma}(\omega,\mathbf{x}-\mathbf{z})\,\tilde{G}(\omega,\mathbf{z}-\mathbf{y})=1\,,
\label{self-energy-equation-momentum}
\end{equation}
where
\begin{equation}
D^\mp=(p_1\mp ip_2)^2 \pm \frac{eB_{\perp}}{\hbar c}(p_1\mp ip_2)(\partial_{p_1}\mp i\partial_{p_2})
+\frac{e^2B^2_{\perp}(\partial_{p_1}\mp i\partial_{p_2})^2}{4(\hbar c)^2}\,.
\label{quadratic}
\end{equation}
In what follows for simplicity we put $\hbar=c=1$.
Equations~(\ref{SD-equation-momentum}) and (\ref{self-energy-equation-momentum}) form a system of two
equations for the translation invariant functions $\tilde{\Sigma}$ and $\tilde{G}$.

Perpendicular magnetic field $B_{\perp}$ enters these equations only through Eq.~(\ref{self-energy-equation-momentum}). Since we seek
the generalized chemical potentials and gaps up to the second order in $B_{\perp}$,
we expand Eq.~(\ref{self-energy-equation-momentum}) up to $B^2_{\perp}$ terms. In fact, we have to
expand only the factor $\mbox{exp}[-ie(\mathbf{x}-\mathbf{y})\mathbf{A(z-y)}]$ because $D^+$ and
$D^-$ given by Eq.~(\ref{quadratic}) are quadratic polynomials in $B_{\perp}$. Expanding the last
term in Eq.~(\ref{self-energy-equation-momentum}) up to the second order in $B_{\perp}$, we find
\begin{eqnarray}
&&\int d^2(\mathbf{x}-\mathbf{y}) e^{-i\mathbf{p}(\mathbf{x}-\mathbf{y})}\int d^2\mathbf{z}\,
e^{-ie(\mathbf{x}-\mathbf{y})\mathbf{A(z-y)}}\tilde{\Sigma}(\omega,\mathbf{x}-\mathbf{z})
\tilde{G}(\omega,\mathbf{z}-\mathbf{y})\nonumber\\
&&\quad\approx \left(1-\frac{ieB_{\perp}\,\partial_{\mathbf{p}}\times\partial_{\mathbf{k}}}{2}
-\frac{e^2B_{\perp}^2}{8}\,(\,\partial_{\mathbf{p}}\times\partial_{\mathbf{k}}\,)^2\right)\,
\tilde{\Sigma}(\omega,\mathbf{p})\,\tilde{G}(\omega,\mathbf{k})|_{\mathbf{k}=\mathbf{p}}\,.
\label{second-term}
\end{eqnarray}
Actually, it is possible to calculate this term in momentum space exactly and it is given by the star product \cite{Douglas} of the translation invariant self-energy and propagator
\begin{equation}
\int d^2(\mathbf{x}-\mathbf{y}) e^{-i\mathbf{p}(\mathbf{x}-\mathbf{y})}\int d^2\mathbf{z}\,
e^{-ie(\mathbf{x}-\mathbf{y})\mathbf{A(z-y)}}\tilde{\Sigma}(\omega,\mathbf{x}-\mathbf{z})
\tilde{G}(\omega,\mathbf{z}-\mathbf{y})=e^{-ieB_{\perp}\,\partial_{\mathbf{p}}
\times\partial_{\mathbf{k}}/2}\tilde{\Sigma}(\omega,\mathbf{p})\, \tilde{G}(\omega,\mathbf{k})|_{\mathbf{k}=\mathbf{p}}\,.
\end{equation}
Thus, Eq.~(\ref{self-energy-equation-momentum}) in the second order in $B_{\perp}$ takes the form
\begin{equation}
\left(\,\omega+ \frac{\hbar^2}{2m_*}
\left( \begin{array}{cc} 0 &
D^- \\
D^+ & 0 \end{array} \right)\,\right)\tilde{G}(\omega,\mathbf{p})
+\left(1-\frac{ieB_{\perp}\,\partial_{\mathbf{p}}\times\partial_{\mathbf{k}}}{2}
-\frac{e^2B_{\perp}^2}{8}\,(\,\partial_{\mathbf{p}}\times\partial_{\mathbf{k}}\,)^2\right)\,
\tilde{\Sigma}(\omega,\mathbf{p})\,\tilde{G}(\omega,\mathbf{k})|_{\mathbf{k}=\mathbf{p}}=1\,.
\label{system-second-equation}
\end{equation}
Equations~(\ref{SD-equation-momentum}) and (\ref{system-second-equation}) define the translation invariant self-energy and propagator up to the second order in $B_\perp$. These equations are
the starting point for the subsequent analysis in Sec.~\ref{5}.
We first rewrite Eq.~\refer{system-second-equation}  as follows:
\be
(\omega+D_0+B_\perp D_1+B_\perp ^2D_2)(\tilde{G}_0+B_\perp \tilde{G}_1+B_\perp ^2\tilde{G}_2)+ (\tilde{\Sigma}_0 +B_\perp \tilde{\Sigma}_1+B_\perp ^2\tilde{\Sigma}_2)(\tilde{G}_0+B_\perp \tilde{G}_1+B_\perp ^2\tilde{G}_2)=1 ,
\label{equation-propagator-expansion}
\ee
where $D_i$ and $\tilde{\Sigma}_i$ can be read off from Eqs.~\refer{quadratic} and~\refer{ansatz}. $\tilde{G}_i~(i=0,1,2)$ represent the full propagators to various orders in $B_{\perp}$. By making
use of Eq.~(\ref{equation-propagator-expansion}), we find
\bea
&&\tilde{G}_0=\frac{1}{\omega+D_0+\tilde{\Sigma}_0}, \label{geo0} \\
&&\tilde{G}_1=-\frac{1}{\omega+D_0+\tilde{\Sigma}_0}(D_1+\tilde{\Sigma}_1)\tilde{G}_0,
\label{geo1}\\
&&\tilde{G}_2=-\frac{1}{\omega+D_0+\tilde{\Sigma}_0}\left[(D_1+\tilde{\Sigma}_1)\tilde{G}_1
+(D_2+\tilde{\Sigma}_2)\tilde{G}_0\right].
\label{geo2}
\eea
Using the obtained expressions for the coefficients of the fermion propagator in expansion
in $B_\perp$, we write down the gap equations  for the studied states.

For the QVH state, we have in the first order in $B_\perp$
\begin{align}
\mu_z^{(1)}-\mub &=\int\frac{d\omega d^2k}{(2\pi)^3}
\frac{\mu_z^{(1)}(\omega^2-E_k^2-m^2)}{(\omega^2+E_k^2+m^2)^2}
\left[V_\text{eff}(\omega,\bk)+\frac{4\pi}{m_*}g_\mu \right],
\\
\mut^{(1)}&=\int\frac{d\omega d^2k}{(2\pi)^3}
\frac{\mut^{(1)}(\omega^2-E_k^2-m^2)-4\chi mE_k}{(\omega^2+E_k^2+m^2)^2}
\left[V_\text{eff}(\omega,\bk)+\frac{4\pi}{m_*}g_{\mut}\right],
\end{align}
where $\chi=e/(2m_*c)$ and
\begin{equation}
g_{\mut}=2g_{z\perp}+g_{0z}-2g_{\perp z}-4g_{\perp\perp}-2g_{\perp0}-7g_{z0}+2g_{0\perp}+g_{zz}.
\end{equation}
In the order $B_\perp^2$, we have
\begin{align}
m^{(2)}&=\int\frac{d\omega d^2k}{(2\pi)^3}\biggl[
\frac{2\chi^2m[9E_k^4-(\omega^2+m^2)^2]}{(\omega^2+E_k^2+m^2)^4}
+\frac{m^{(2)}(E_k^2+\omega^2-m^2)}{(\omega^2+E_k^2+m^2)^2}
\nonumber \\
&\quad+\frac{(E_k^2+m^2-3\omega^2)\{m[(\mu_z^{(1)})^2+(\mut^{(1)})^2]+4\chi\mut^{(1)} E_k\}}{(\omega^2+E_k^2+m^2)^3}\biggr]
\left[V_\text{eff}(\omega,\bk)+\frac{4\pi}{m_*}g_m\right],
\\
\Delta_z^{(2)}&=\int\frac{d\omega d^2k}{(2\pi)^3}\biggl[
\frac{\Delta_z^{(2)}(E_k^2+\omega^2-m^2)}{(\omega^2+E_k^2+m^2)^2}
+\frac{2\mu_z^{(1)}(m\mut^{(1)}+2\chi E_k)(E_k^2+m^2-3\omega^2)}{(\omega^2+E_k^2+m^2)^3}\biggr]
\left[V_\text{eff}(\omega,\bk)+\frac{4\pi}{m_*}g_{\Delta_z}\right]
\end{align}
[for the gap equation in the zeroth order in $B_\perp$, see Eq.~(\ref{qvhgeq3})].
For the LAF state, the zeroth order in $B_\perp$ gap equation is given by Eq.~(\ref{gengeq3}) with $\delta=m_x$, the $\mathcal O(B_\perp)$ gap
equation reads
\begin{align}
\mu_z^{(1)}-\mub&=-\frac{\mu_z^{(1)}}{2}\int\frac{d\omega d^2k}{(2\pi)^3}\sum_{\lambda=\pm}
\biggl(1+\frac{\lambda(\omega^2+m_x^2)}{m m_x}\biggr)\frac{1}{\omega^2+E_k^2+m_\lambda^2}
\left[V_\text{eff}(\omega,\bk)+\frac{4\pi}{m_*}g_{\mu}\right],
\\
\mut_\pm&=\int\frac{d\omega d^2k}{(2\pi)^3}\sum_{\lambda=\pm}\left[\frac{1\pm\lambda}{2}
V_\text{eff}(\omega,k)+\frac{2\pi}{m_*}(g_{\mut}\pm\lambda g_{\mut_z})\right]
\biggl[\frac{2(\mut_\lambda\omega^2-2\chi m_\lambda E_k)}{(\omega^2+E_k^2+m_\lambda^2)^2}
-\frac{\mut_\lambda}{\omega^2+E_k^2+m_\lambda^2}\biggr],
\end{align}
and in the order $B_\perp^2$
\begin{align}
m^{(2)}_\pm&=\int\frac{d\omega d^2k}{(2\pi)^3}\sum_{\lambda=\pm}\Biggl\{
\left[\frac{1\pm\lambda}{2}V_\text{eff}(\omega,\bk)+\frac{2\pi}{m_*}(g_m\pm\lambda g_{m_z})\right]
\biggl[\frac{16\chi^2m_\lambda E_k^4}{(\omega^2+E_k^2+m_\lambda^2)^4}
\nonumber \\
&\quad+\frac{4[\chi^2m_\lambda E_k^2+\mut_\lambda(m_\lambda\mut_\lambda+4\chi E_k)(E_k^2+m_\lambda^2)]}
{(\omega^2+E_k^2+m_\lambda^2)^3}
-\frac{m_\lambda(2m_\lambda m^{(2)}_\lambda+3\mut_\lambda^2+2\chi^2)+12\chi\mut_\lambda E_k}
{(\omega^2+E_k^2+m_\lambda^2)^2}
\nonumber \\
&\quad+\frac{4m_x^2m^2m^{(2)}_\lambda-(\mu_z^{(1)})^2[m_\lambda(E_k^2+m^2)-2\lambda m_xm^2]}
{4m_x^2m^2(\omega^2+E_k^2+m_\lambda^2)}
-\frac{(\mu_z^{(1)})^2m_\lambda(E_k^2+m_\lambda m)}{\lambda m_xm(\omega^2+E_k^2+m_\lambda^2)^2}\biggr]
\nonumber \\
&\quad+\left[\frac{1\mp\lambda}{2}V_\text{eff}(\omega,\bk)+\frac{2\pi}{m_*}(g_m\mp\lambda g_{m_z})\right]
\frac{(\mu_z^{(1)})^2[m_\lambda(Ek^2+m^2)-2\lambda m_x E_k^2]}{4m_x^2m^2(\omega^2+E_k^2+m_\lambda^2)}\Biggr\},
\\
\Delta^{(2)}_z&=\int\frac{d\omega d^2k}{(2\pi)^3}\sum_{\lambda=\pm}\frac{\lambda}{2m_xm}
\biggl[\frac{2\mu_z^{(1)}[m(\mut_\lambda\omega^2-2\chi m_\lambda E_k)-2\chi E_k^3]}
{(\omega^2+E_k^2+m_\lambda^2)^2}
-\frac{\mu_z^{(1)}\mut_x^{(1)}(E_k^2+m^2)}{m_x(\omega^2+E_k^2+m_\lambda^2)}
\nonumber \\
&\quad+\frac{\mu_z^{(1)}[m(\mut^{(1)}-2\mut_\lambda)+4\chi E_k]+m_\lambda m\Delta^{(2)}_z}{\omega^2+E_k^2+m_\lambda^2}\biggr]
\left[V_\text{eff}(\omega,\bk)+\frac{4\pi}{m_*}g_{\Delta_z}\right],
\end{align}
where $m_\pm=m\pm m_x$, $\mut_\pm=\mut^{(1)}\pm\mut_x^{(1)}$, and $m_\pm^{(2)}=m^{(2)}\pm m_x^{(2)}$.
The energy density~\refer{energy-density-momentum} expanded up to the second order in $B_\perp $
takes the form
\bea
{\cal E}&=&\frac{i}{2}\int \frac{d\omega d^2p}{(2\pi)^3}\,\mbox{tr}\,
\left\{\left[\,\left(\,-\omega-m_0\eta_3\tau_3 + D_0\right)+\left(-\mu_BB_{\perp}\sigma_3+B_\perp D_1\right)+B_\perp ^2D_2\right] \left(\tilde{G}_0+B_\perp \tilde{G}_1+B_\perp ^2\tilde{G}_2\right)\,\right\}\nonumber\\
&&- (\mu_z,\tilde{\mu},\tilde{\mu}_z,m^{(2)},m^{(2)}_z,B_\perp \to 0)\nonumber\\
&\equiv& \frac{i}{2}\int \frac{d\omega d^2p}{(2\pi)^3}\,\mbox{tr}\,
\left\{F_0\tilde{G}_0+B_\perp (F_0\tilde{G}_1+F_1\tilde{G}_0)+B_\perp ^2(F_0\tilde{G}_2+F_1\tilde{G}_1+F_2\tilde{G}_0)\right\}\nonumber\\
&&- (\mu_z,\tilde{\mu},\tilde{\mu}_z,m^{(2)},m^{(2)}_z,B_\perp \to 0),
\label{edb2}
\eea
where
\be
F_0=\left(-\omega-m_0\eta_3\tau_3 + D_0\right),\quad
F_1=\mu_B \sigma_3+D_1,\quad
F_2=D_2.
\ee
The gap equations in a weak perpendicular magnetic field derived in this Appendix are
solved numerically, and the results are presented in Sec.~\ref{5}.


\begin{thebibliography}{99}

\bibitem{McC} E. McCann and V.~I.~Fal'ko, Phys. Rev. Lett. {\bf 96}, 086805 (2006).

\bibitem{Min} F.~Zhang, H.~Min, M.~Polini, and A.~H.~MacDonald, Phys. Rev. B {\bf 81},
041402 (2010).

\bibitem{GGMSh} E.~V.~Gorbar, V.~P.~Gusynin, V.~A.~Miransky, and I.~A.~Shovkovy, Phys. Rev.
B {\bf 66}, 045108 (2002).

\bibitem{Khveshchenko} D.~V.~Khveshchenko and H.~Leal, Nucl. Phys. B {\bf687}, 323 (2004).

\bibitem{Gamayun} O.~V.~Gamayun, E.~V.~Gorbar, and V.~P.~Gusynin, Phys. Rev. B {\bf 81},
075429 (2010).

\bibitem{Fertig} J.~Wang, H.~A.~Fertig, and G.~Murthy, Phys. Rev. Lett. {\bf104}, 186401 (2010).

\bibitem{Gonzalez} J.~Gonz\'alez, Phys. Rev. B {\bf85}, 085420 (2012).

\bibitem{Levitov}R.~Nandkishore and L.~Levitov, Phys. Rev. Lett. {\bf104}, 156803 (2010).

\bibitem{footnote} The different role of interactions for semimetals with linear and quadratic
electron dispersion was recognized in an early work:
A.A.~Abrikosov and S.D.~Beneslavskii, Zh. Eksp. Teor. Fiz. {\bf59}, 1280 (1970)
[Sov. Phys. JETP {\bf32}, 4 (1971)]; J. Low Temp. Phys. {\bf5}, 141 (1971).


\bibitem{Mayorov} A.~S.~Mayorov, D.~C.~Elias, I.~S.~Mukhin, S.~V.~Morozov,
L.~A.~Ponomarenko, K.~S.~Novoselov, A.~K.~Geim, and R.~V.~Gorbachev, Nano Lett.
{\bf 12}, 4629 (2012).

\bibitem{Weitz} R.~T.~Weitz, M.~T.~Allen, B.~E.~Feldman, J.~Martin, and A.~Yacoby,
Science {\bf 330}, 812 (2010).

\bibitem{Freitag} F.~Freitag, J.~Trbovic, M.~Weiss, and C.~Sch\"onenberger, Phys. Rev.
Lett. {\bf 108}, 076602 (2012).

\bibitem{Velasco} J.~Velasco Jr., L.~Jing, W.~Bao, Y.~Lee, P.~Kratz, V.~Aji, M.~Bockrath,
C.~N.~Lau, C.~Varma, R.~Stillwell, D.~Smirnov, F.~Zhang, J.~Jung, and A.~H.~MacDonald,
Nat. Nanotechnol. {\bf 7}, 156 (2012).

\bibitem{Bao} W.~Bao, J.~Velasco, F.~Zhang, L.~Jing, B.~Standley, D.~Smirnov,
M.~Bockrath, A.~H.~MacDonald, and C.~N.~Lau, Proc. Natl. Acad. Sci. USA {\bf 109},
10802 (2012).

\bibitem{catalysis} V.~P.~Gusynin, V.~A.~Miransky, and I.~A.~Shovkovy,
Phys. Rev. Lett. {\bf 73}, 3499 (1994).

\bibitem{MC} V.~P.~Gusynin, V.~A.~Miransky, S.~G.~Sharapov, and I.~A.~Shovkovy,
Phys. Rev. B {\bf 74}, 195429 (2006); I.~F.~Herbut, {\it ibid.} {\bf 75}, 165411
(2007); J.-N.~Fuchs and P.~Lederer, Phys. Rev. Lett. {\bf 98}, 016803 (2007);
M.~Ezawa, J. Phys. Soc. Jpn. {\bf 76}, 094701 (2007).

\bibitem{Levitov-anomalous} R.~Nandkishore and L.~Levitov, Phys. Rev. B {\bf 82},
115124 (2010).

\bibitem{Fzhang} F.~Zhang, J.~Jung, G.~A.~Fiete, Q.~Niu, and A.~H.~MacDonald, Phys. Rev.
Lett. {\bf 106}, 156801 (2011).

\bibitem{MacDonald} F.~Zhang and A.~H.~MacDonald, Phys. Rev. Lett. {\bf 108}, 186804 (2012).

\bibitem{Haldane} F.~D.~M.~Haldane, Phys. Rev. Lett. {\bf 61}, 2015 (1988).

\bibitem{Martin} J.~Martin, B.~E.~Feldman, R.~T.~Weitz, M.~T.~Allen, and A.~Yacoby,
Phys. Rev. Lett. {\bf 105}, 256806 (2010).

\bibitem{Feldman} B.~E.~Feldman, J.~Martin, and A.~Yacoby, Nat. Phys. {\bf 5}, 889 (2009).

\bibitem{Zhao} Y.~Zhao, P.~Cadden-Zimansky, Z.~Jiang, and P.~Kim, Phys. Rev. Lett. {\bf 104},
066801 (2010).

\bibitem{Kim} S.~Kim, K.~Lee, and E.~Tutuc, Phys. Rev. Lett. {\bf 107}, 016803 (2011).

\bibitem{Elferen} H.~J.~van Elferen, A.~Veligura, E.~V.~Kurganova, U.~Zeitler, J.~C.~Maan,
N.~Tombros, I.~J.~Vera-Marun, and B.~J.~van Wees, Phys. Rev. B {\bf 85}, 115408 (2012).

\bibitem{Hunt}B.~M.~Hunt, J.~I.~A.~Li, A.~A.~Zibrov, L.~Wang, T.~Taniguchi, K.~Watanabe, J.~Hone,
C.~R.~Dean, M.~Zaletel, R.~C.~Ashoori, and A.~F.~Young, arXiv:1607.06461.

\bibitem{Kharitonov1} M.~Kharitonov, Phys. Rev. Lett. {\bf 109}, 046803 (2012).

\bibitem{Barlas} Y.~Barlas, R.~C\^ot\'e, K.~Nomura, and A.~H.~MacDonald,
Phys. Rev. Lett. {\bf 101}, 097601 (2008).

\bibitem{Abergel} D.~S.~L.~Abergel and T.~Chakraborty, Phys. Rev. Lett. {\bf102}, 056807 (2009).

\bibitem{Shizuya} K.~Shizuya,  Phys. Rev. {\bf 79}, 165402 (2009).

\bibitem{Nakamura} M.~Nakamura, E.~V.~Castro, and B.~D\'ora, Phys. Rev. Lett. {\bf 103},
266804 (2009).

\bibitem{GGM1} E.~V.~Gorbar, V.~P.~Gusynin, and V.~A.~Miransky, Pis'ma Zh. Eksp. Teor. Fiz. {\bf 91},
334 (2010) [JETP Lett. {\bf91}, 314 (2010)];
Phys. Rev. B {\bf 81}, 155451 (2010).

\bibitem{Nandkishore1} R.~Nandkishore and L.~Levitov, Phys. Scr. {\bf T146}, 014011 (2012).

\bibitem{Toke} C.~T\H{o}ke and V.~I.~Fal'ko, Phys. Rev. B {\bf83}, 115455 (2011).

\bibitem{GGJM} E.~V.~Gorbar, V.~P.~Gusynin, J.~Jia, and V.~A.~Miransky,
Phys. Rev. B {\bf 84}, 235449 (2011).

\bibitem{GGMSh-mixing} E.~V.~Gorbar, V.~P.~Gusynin, V.~A.~Miransky, and I.~A.~Shovkovy,
Phys. Rev. B {\bf 85}, 235460 (2012).

\bibitem{Kharitonov} M. Kharitonov, Phys. Rev. B {\bf 86}, 075450 (2012).

\bibitem{Kharitonov-AF} M. Kharitonov, Phys. Rev. B {\bf 86}, 195435 (2012).

\bibitem{Lemonik}
Y.~Lemonik, I.~L.~Aleiner, C.~Toke, and V.~I.~Fal'ko, Phys. Rev. B {\bf 82}, 201408(R) (2010);
Y.~Lemonik, I.~Aleiner, and V.~I.~Fal'ko, {\it ibid.} {\bf 85}, 245451 (2012).

\bibitem{Vafek}
O.~Vafek, Phys. Rev. B {\bf 82}, 205106 (2010);
V.~Cvetkovic, R.~E.~Throckmorton, and O.~Vafek, {\it ibid.} {\bf 86}, 075467 (2012);
R.~E.~Throckmorton and O.~Vafek, {\it ibid.} {\bf 86}, 115447 (2012).

\bibitem{Kharitonov-mono} M.~Kharitonov, Phys. Rev. B {\bf85}, 155439 (2012)

\bibitem{Maher} P.~Maher, C.~R.~Dean, A.~F.~Young, T.~Tanugichi, K.~Watanabe, K.~L.~Shepard,
J.~Hone, and P.~Kim, Nat. Phys. {\bf9}, 154 (2013).

\bibitem{Freitag2} F.~Freitag, M.~Weiss, R.~Maurand, J.~Trbovic, and C.~Sch\"onenberger,
Phys. Rev. B {\bf 87}, 161402(R) (2013).

\bibitem{Pershoguba} S.~S.~Pershoguba and V.~M.~Yakovenko, Phys. Rev. B {\bf82}, 205408 (2010).

\bibitem{Roy3} B.~Roy and K.~Yang, Phys. Rev. B {\bf88}, 241107(R) (2013).

\bibitem{Peeters} M.~Van der Donck, F.~M.~Peeters, and B.~Van Duppen, Phys. Rev. B {\bf93}, 115423 (2016).

\bibitem{Falko} N.~Kheirabadi, E.~McCann, and V.~I.~Fal'ko, Phys. Rev. B {\bf94}, 165404 (2016).

\bibitem{Roy} B.~Roy, M.P.~Kennett, and S.~Das Sarma, Phys. Rev. B {\bf90}, 201409(R) (2014).

\bibitem{Roy2} B.~Roy, Phys. Rev. B {\bf89}, 201401(R) (2014).

\bibitem{Klimenko} K.~G.~Klimenko and R.~N.~Zhokhov, Phys. Rev. D {\bf88}, 105015 (2013).

\bibitem{Pezzini} S.~Pezzini, C.~Cobaleda, B.~A.~Piot, V.~Bellani, and E.~Diez, Phys. Rev. B
{\bf 90}, 121404(R) (2014).

\bibitem{Guinea} P.~San-Jose, J.~L.~Lado, R.~Aguado, F.~Guinea, and J.~Fernandez-Rossier, 
Phys. Rev. X {\bf 5}, 041042 (2015).

\bibitem{Douglas} M.~R.~Douglas and N.~A.~Nekrasov, Rev. Mod. Phys. {\bf 73}, 977 (2001).

\end{thebibliography}
\end{document}